\documentclass[twocolumn]{emulateapj}
\usepackage{graphicx}
\usepackage{amssymb,amsmath}

\slugcomment{Accepted for publication in the Astrophysical Journal, September 1, 2011}

\begin{document}
\shorttitle{Likelihood}
\shortauthors{Kirkpatrick, et. al}
\title{\vspace{-0.4cm} A Simple Likelihood Method for Quasar Target Selection}

\author{
Jessica A. Kirkpatrick\altaffilmark{1}$^,$\altaffilmark{2},
David J. Schlegel\altaffilmark{2},
Nicholas P. Ross\altaffilmark{2},
Adam D. Myers\altaffilmark{3}$^,$\altaffilmark{4}$^,$\altaffilmark{5},\\
Joseph F. Hennawi\altaffilmark{4},
Erin S. Sheldon\altaffilmark{6},
Donald P. Schneider\altaffilmark{7},
Benjamin A. Weaver\altaffilmark{8}
}

\altaffiltext{1}{The University of California, Department of Physics, Berkeley, CA 94720, USA; kirkpatrick@berkeley.edu}
\altaffiltext{2}{Lawrence Berkeley National Laboratory, 1 Cyclotron Rd, Berkeley, CA 92420, USA}
\altaffiltext{3}{University of Illinois, 1002 West Green Street, Urbana, IL 61801, USA}
\altaffiltext{4}{Max-Planck-Institut fŸr Astronomie, Kšnigstuhl 17, 69117 Heidelberg, Germany}
\altaffiltext{5}{University of Wyoming, Department of Physics \& Astronomy, Laramie, WY 82071, USA}
\altaffiltext{6}{Brookhaven National Laboratory, Physics Department, Mail Stop: 510A, Upton, NY 11973-5000}
\altaffiltext{7}{The Pennsylvania State University, Department of Astronomy \& Astrophysics, University Park, PA 16802, USA}
\altaffiltext{8}{New York University, Center for Cosmology and Particle Physics, New York, NY 10003 USA}

\begin{abstract}
We present a new method for quasar target selection using photometric fluxes and a Bayesian probabilistic approach.  For our purposes we target quasars using Sloan Digital Sky Survey (SDSS) photometry to a magnitude limit of $g=22$.  The efficiency and completeness of this technique is measured using the Baryon Oscillation Spectroscopic Survey (BOSS) data, taken in 2010.  This technique was used for the uniformly selected (CORE) sample of targets in BOSS year one spectroscopy to be realized in the 9th SDSS data release.  When targeting at a density of 40 objects per sq-deg (the BOSS quasar targeting density) the efficiency of this technique in recovering $z>2.2$ quasars is 40\%.  The completeness compared to all quasars identified in BOSS data is 65\%.  This paper also describes possible extensions and improvements for this technique.
\end{abstract}

\maketitle

\section{Introduction}
\label{sec:Introduction}

The SDSS-III: Baryon Oscillation Spectroscopic Survey \citep[BOSS;][]{Eisenstein11} is specifically targeting $z > 2.2$ QSOs in order to observe 150,000 Lyman-$\alpha$ forest (Ly$\alpha$F) lines of sight. The key aim of BOSS is to measure the absolute cosmic distance scale and expansion rate with percent-level precision at three distinct cosmological epochs: redshifts $z = 0.3, 0.6$ using luminous red galaxies (LRGs) and $z\sim2.5$ using the Ly$\alpha$F as the density tracer.  For both the galaxy and Ly$\alpha$F samples, the primary distance measure is the baryon acoustic oscillation (BAO) technique \citep{Schlegel2007, Schlegel2009,Slosar2011}.  BOSS dedicates 40 fibers deg$^{-2}$ to QSO target selection for measuring the BAO signal.

Previous quasar surveys, such as the Sloan Digital Sky Survey \citep[SDSS;][]{Schneider10} and the Ango-Australian Telescope Two-Degree Field (2dF) QSO Redshift Survey \citep[2QZ;][]{Croom04}, have historically performed quasar target selection by searching for relatively bright quasars ($i<19.1$, $z<3$ objects for SDSS). However, previous methods, such as the traditional ``UV Excess'' \citep[UVX; selecting star-like objects with unusually blue broadband colors,][]{Sandage1965}, ``color-boxes'' \citep{Croom2009} and ``Kernel Density Estimators'' \citep{Richards2004} begin to fail at fainter magnitudes because photometric errors broaden the stellar locus, leading to potential incompleteness and inefficiency in target selection. This motivated our development of a selection technique which better handles the photometric flux errors as one approaches the flux limit. 

Furthermore, at redshifts between $z=2.5$ and $z=3.0$, broad-band optical color selections fail, since the colors of these quasars are similar to those of stars (in particular early A and F stars, \citealt{Fan99}, \citealt{Richards02}) as the quasars ``pass over the stellar locus'' when the photometric colors are the same as the stellar colors (see Section~\ref{subsec:performance} for more details).  Simultaneously, quasars become much fainter e.g., an $M_{g}=-23$ quasar at $z=2$, has $g$-band $\sim21.7$, which is close to the SDSS single-epoch magnitude limit. 

The BOSS Ly$\alpha$F/Quasar Survey will target objects thought to be $z>2.2$ quasars to perform a Ly$\alpha$F BAO measurement. Since the foreground Ly$\alpha$F is independent of the intrinsic properties of the background quasar, there is freedom to use multiple selection methods without biasing the BAO results. The methods used for BOSS targeting include the ``Kernel Density Estimator'' \citep[KDE;][]{Richards2004}, an ``Extreme-Deconvolution'' method \citep[XDQSO;][]{Bovy2011}, and a Neural Network method \citep[NN;][]{Yeche10}. The BOSS QSO target selection used for the first year of observations \citep{Ross11} combines all these different methods, including the Likelihood method described in this paper, with different photometric catalogs such as SDSS \citep{York2000}, UKIDSS \citep{Lawrence2007}, GALEX \citep{Martin2005} and quasars found using their flux time-variability information \citep{PD11}.

In this paper we describe a new method for quasar target selection. Our method models data in 5-filter flux space, then calculates likelihood estimates that a given object is a $z>2.2$ quasar.  Because a given survey has a finite number of spectroscopic fibers (observing time allocation) to dedicate towards quasar targeting, this method attempts to prioritize selection by calculating a probability that a potential target is a quasar based on these likelihood calculations. Targets are ranked by likelihood probability.  This method differs from KDE in that it incorporates the photometric errors for each object into the likelihood calculations; also KDE only imposes a single magnitude prior and color-distribution, whereas we model the QSO density as a function of magnitude and evolution of color distribution. 

The layout of this paper is as follows.  Section (\ref{sec:Method}) describes the method used to calculate the likelihoods, and training catalogs that are generated and used for Likelihood target selection. In Section (\ref{sec:commissioning}) we give an overview of the BOSS Data and the performance of the Likelihood method using this data. In Section (\ref{sec:conclusions}) we discuss testing and optimization of the method, as well as future work and possible improvements.  We use the terms ``quasar'' and ``QSO'' interchangeably to refer to quasi-stellar, type-I broad line objects.  All Right Ascension (RA) and Declinations (Dec) discussed are J2000.

Our Likelihood method was used for the uniformly selected sample (which we refer to hereafter as ``CORE'') in the first year of the BOSS QSO target selection \citep{Ross11} and it is our intention to release our calculated likelihood probabilities as a data product in the future Data Releases of the BOSS (the first such event is SDSS Data Release 9). 

\section{Method and Catalog Generation}
\label{sec:Method}

\subsection{Likelihood Method}
\label{subsec:likelihood}

Recent work has approached target selection within a Bayesian statistical framework over more traditional color-box approaches \citep{Richards2004, Bovy2011, Mortlock11}. Spectroscopic target selection can be viewed as a classification problem.  Given a set of photometric target objects ($O$) with attributes ($\mathbf{a}$) and a discrete set of astronomical object classes, one would like to assign a target to a particular class.  For our purposes we are simply interested in the question: ``Is the object a quasar?''  Thus we have two classes: quasar ($QSO$) and non-quasar (i.e., all other observable objects: stars + galaxies + anything else) hereby referred to as Everything Else ($EE$).

The probability that an object $O$ is a quasar (in class $QSO$) given a vector of object attributes $\mathbf{a}$, is provided by Bayes' theorem \citep{Sivia2006}:
\begin{equation}
\label{probability1}
\fontsize{9}{13}\mathcal{P}(O\in QSO \mid \mathbf{a}\,) = \frac{\mathcal{P}(\,\mathbf{a} \mid O \in QSO) \;\mathcal{P}(O \in QSO)}{\mathcal{P}(\mathbf{a})}
\end{equation}
where $\mathcal{P}(\,\mathbf{a} \mid O \in QSO)$ is the conditional probability that given attributes $\mathbf{a}$, object $O$ is a quasar; $\mathcal{P}(O~\in~QSO)$ is the prior probability that $O$ is a quasar (prior in the sense that it does not take into account any information about the object attributes); $\mathcal{P}(\mathbf{a})$ is the marginal probability of an object with attributes $\mathbf{a}$ occurring at all, and acts as a normalizing constant.  In our case: 
\begin{equation}
\fontsize{9}{13}
\label{probdenominator}
\begin{aligned}[]
\mathcal{P}(\mathbf{a}) =\;&\mathcal{P}(\,\mathbf{a} \mid O \in QSO) \;\mathcal{P}(O \in QSO) \\
+\;&\mathcal{P}(\,\mathbf{a} \mid O \in EE) \;\mathcal{P}(O \in EE)
\end{aligned}
\end{equation}
because $QSO \cup EE$ contain all possible classifications (or outcomes) for object $O$.

We used the term ``likelihood'' to denote the conditional probabilities $\mathcal{P}(\,\mathbf{a} \mid O \in QSO)$ and $\mathcal{P}(\,\mathbf{a}~\mid~O~\in~EE)$ in Eqns.\;(\ref{probability1}) and (\ref{probdenominator}).  In the case where the attributes of a target object are measured with a significant amount of uncertainty, one can imagine $\mathbf{a}$ is a noisy measurement of an underlying true attribute vector $\mathbf{a}'$. We can then calculate the likelihood by marginalizing over all possible of values of $\mathbf{a}'$:\;\footnote[1]{In the derivation of Eqn.\;(\ref{likelihood}), we assume the noisy observation $\mathbf{a}$ is independent of the classification of $O$ given $\mathbf{a}'$, therefore:\\*\centering$\mathcal{P}(\,\mathbf{a}~\mid~\mathbf{a}', O \in QSO) = \mathcal{P}(\,\mathbf{a} \mid \mathbf{a}')$.}
\begin{equation}
\fontsize{9}{13}
\begin{aligned}[]
\label{likelihood}
\mathcal{P}(\,\mathbf{a} &\mid O \in QSO) = \int \mathcal{P}(\,\mathbf{a},\mathbf{a}' \mid O \in QSO) \;d\mathbf{a}' \\
&= \int \mathcal{P}(\,\mathbf{a} \mid \mathbf{a}', O \in QSO) \; \mathcal{P}(\,\mathbf{a}' \mid O \in QSO) \;d\mathbf{a}' \\
&= \int \mathcal{P}(\,\mathbf{a} \mid \mathbf{a}') \; \mathcal{P}(\,\mathbf{a}' \mid O \in QSO) \;d\mathbf{a}'\;.
\end{aligned}
\end{equation}

For our purposes, $\mathcal{P}(\,\mathbf{a}' \mid O \in QSO)$ is just the empirical distribution observed in a discrete set of high signal-to-noise objects which are already classified to be either quasars or non-quasars.  The attributes are the photometric fluxes ($f$) in the five SDSS color filters ($\mathbf{f}=\{u,~g,~r,~i,~z\}$) and are independent of each other.  Because the empirical distribution places a $\delta$-function at each training example in Eqn.\;(\ref{likelihood}), the integral becomes a sum over all objects ($O'$) with attributes $\mathbf{a}'$ in the training sets:  
\begin{equation}
\label{likelihoodsum}
\fontsize{9}{13}\int \mathcal{P}(\,\mathbf{a} \mid \mathbf{a}') \; \mathcal{P}(\,\mathbf{a}' \mid O \in QSO) \;d\mathbf{a}' = \displaystyle\sum_{O'}\;\mathcal{P}(\,\mathbf{a} \mid \mathbf{a}')\;.
\end{equation}

Like other recent publications \citep{Bovy2011, Mortlock11}, we use a Gaussian distribution, $\mathcal{P}(\,\mathbf{a}~\mid~\mathbf{a}')$, for the uncertainties of the attributes.  Thus a Gaussian distribution is used for the errors ($\sigma_{f}$) in the object fluxes, and fluxes $f$ and $f^{O'}$ for one of the target object attributes ($\mathbf{a}$) and training object attributes ($\mathbf{a}'$) respectively.  The likelihood ($\mathcal{L}$) for a single flux $f$ then becomes:
\begin{equation}
\fontsize{9}{13}
\label{like1}
\begin{aligned}[]
\mathcal{L} &= \mathcal{P}(\,f \mid O \in QSO\mbox{ or }EE) \\
&\simeq \displaystyle\sum_{O'}\sqrt{\frac{1}{2\pi\sigma_{f}^2}} \exp{\left[-\frac{[f-f^{O'}]^2}{2\sigma_{f}^2} \right]}\;.
\end{aligned}
\end{equation}
When we consider all five SDSS fluxes, there is a multiplicative sum over these attributes and Eqn.\;(\ref{like1}) becomes:
\begin{equation}
\fontsize{9}{13}
\label{like2}
\begin{aligned}[]
\mathcal{L} &= \mathcal{P}(\{u,g,r,i,z\} \mid O \in QSO\mbox{ or }EE) \\
&\simeq \displaystyle\sum_{O'}\displaystyle\prod_{f=u,g,r,i,z}\sqrt{\frac{1}{2\pi\sigma_{f}^2}} \exp{\left[-\frac{[f-f^{O'}]^2}{2\sigma_{f}^2} \right]}\;\;.
\end{aligned}
\end{equation}
Note that the above equations become equalities when the training catalogs completely represent the object flux-space.  For our target object fluxes ($f$), we used SDSS photometric PSF fluxes from the SkyServer (http://www.sdss3.org/dr8/) under the standard SDSS data releases. 

All fluxes and their errors are corrected for Galactic extinction in the SDSS filters using the prescription in \cite{SFD98}.  Because the sum is done in flux space rather than color space, object errors are independent.  Also, our method preserves the luminosity function information, whereas the absolute flux information is lost when using colors.  Our training catalogs use stacked (co-added) fluxes (see Sections~\ref{subsec:QSOCatalog} and \ref{subsec:EECatalog}), whereas the targets are single epoch fluxes (see Section~\ref{subsec:performance}).  Therefore in the above equations, the errors in the catalog fluxes (${f^{O'}}$) are ignored because the signal-to-noise ratio of the catalog fluxes are much greater than the signal-to-noise ratio of our potential target fluxes (${f}$).  

The QSO likelihood can separated into redshift bins $(\Delta z)$ so that we can tune Eqn.\;(\ref{probability1}) to a desired target redshift range.  This simply requires having redshifts for the objects ($O'$) in the QSO training catalog and subdividing this data into redshift bins.  We used a width $\Delta z = 0.1$ (e.g. $0.5\rightarrow0.6$, $0.6\rightarrow0.7$ ... $4.9\rightarrow5.0$).  This results in the following final equations for the $QSO$ and $EE$ likelihoods:
\begin{equation}
\fontsize{9}{13}
\begin{aligned}
\label{likeQSO}\mathcal{L}_{QSO}&(\Delta z) = \mathcal{P}(\,\mathbf{f} \mid O \in QSO(\Delta z)) \\
&\simeq \!\!\!\!\!\!\!\!\displaystyle\sum_{{O' \in QSO(\Delta z)}} \;  \displaystyle\prod_{f=\mathbf{f}} \sqrt{\frac{1}{2\pi\sigma_{f}^2}} \exp{\left[-\frac{(f - f^{O'})^2}{2\sigma_{f}^2} \right]}\,,
\end{aligned}
\end{equation}
\begin{equation}
\fontsize{9}{13}
\begin{aligned}
\label{likeEE}
\mathcal{L}_{EE} &= \mathcal{P}(\,\mathbf{f} \mid O \in EE) \\
&\simeq \displaystyle\sum_{O' \in EE} \; \displaystyle\prod_{f=\mathbf{f}} \sqrt{\frac{1}{2\pi\sigma_{f}^2}} \exp{\left[-\frac {(f-f^{O'})^2}{2\sigma_{f}^2} \right]} \,.
\end{aligned}
\end{equation}
The Gaussian normalizations add a multiplicative constant to each likelihood ($\mathcal{L}$), which is the same for both Eq.\;(\ref{likeQSO}) and Eq.\;(\ref{likeEE}) for a given target and cancel when calculating the probabilities in Eqn.\;(\ref{probability1}).  

The prior probabilities in Eqn.\;(\ref{probability1}) and Eqn.\;(\ref{probdenominator}) are the relative surface densities of quasars and everything else on the sky, and thus normalize Eqn.\;(\ref{likeQSO}) and Eqn.\;(\ref{likeEE}).  We do this by defining the prior probabilities to be the inverse of the effective sky area ($A$) of the $QSO$ and $EE$ catalogs:
\begin{equation}
\label{priorQSOEE}
\fontsize{9}{13}\mathcal{P}(O \in QSO) = \frac{1}{A_{QSO}}\mbox{ , }\;\;\;\;\mathcal{P}(O \in EE) = \frac{1}{A_{EE}}\;.
\end{equation}
By inserting Eqn.\;(\ref{likeEE}), Eqn.\;(\ref{likeQSO}), and Eqn.\;(\ref{priorQSOEE}) into  Eqn.\;(\ref{probability1}) we can get probability that a single potential target object ($O$) is a quasar ($QSO$) in a target redshift range ($\Delta z_{\mathrm{target}}$):
\begin{equation}
\label{probability}
\fontsize{9}{13}\mathcal{P}(O\in QSO(\Delta z_{\mathrm{target}}) \mid \mathbf{f}\;) \simeq \frac{\displaystyle\sum_{{\Delta z}_{\mathrm{target}}}\left(\frac{\mathcal{L}_{QSO}(\Delta z)}{A_{QSO}}\right)}{\displaystyle\frac{\mathcal{L}_{EE}}{A_{EE}}+\displaystyle\sum_{{\Delta z}_{\mathrm{all}}}\left(\frac{\mathcal{L}_{QSO}(\Delta z)}{A_{QSO}}\right)}\;\;.
\end{equation}
In the numerator, $\mathcal{L}_{QSO}(\Delta z)$ is summed over the desired quasar target redshift range ($\Delta z_{\mathrm{target}}$), whereas the denominator contains all objects in both catalogs summed over the entire redshift range ($\Delta z_{\mathrm{all}}$).  This probability is exact in the limit of perfect training catalogs (infinite objects and zero errors).  A probability is calculated for every potential target using the full $QSO$ and $EE$ catalogs.

\subsection{Imaging Data}
BOSS uses the same imaging data as that of the original SDSS-I/II survey \citep{York2000}, with an extension in the South Galactic Cap (SGC). These data were gathered using a dedicated 2.5 m wide-field telescope \citep{Gunn2006} to collect light for a camera with 30 2k$\times$2k CCDs \citep{Gunn1998} over five broad bands - {\it ugriz} \citep{Fukugita1996}; this camera has imaged 14,555 deg$^{2}$ of the sky, including \hbox{7,500 deg$^{2}$} in the North Galactic Cap (NGC) and \hbox{3,100 deg$^{2}$} in the SGC \citep{SDSSDR82011}. The imaging data were taken on dark photometric nights of good seeing \citep{Hogg01}, and objects were detected and their properties were measured \citep{Lupton01, Stoughton02} and calibrated photometrically \citep{Smith02, Ivezic04, Tucker06, Padmanabhan08a}, and astrometrically \citep{Pier03}.

\citet{Padmanabhan08a} present an algorithm which uses overlaps between SDSS imaging scans to photometrically calibrate the SDSS imaging data.  BOSS target selection uses these ``ubercalibrated'' data from the SDSS Data Release Eight (DR8) database \citep{SDSSDR82011}.  The $2.5^\circ$ stripe along the celestial equator in the Southern Galactic Cap, commonly referred to as ``Stripe 82'' was imaged multiple times, for up to 80 epochs spanning a 10-year baseline \citep{Abazajian09}. A coaddition of these data \citep{Adelman-McCarthy08} goes roughly two magnitudes fainter than the single-scan images which make up the bulk of the SDSS imaging data.

\subsection{QSO Catalog}
\label{subsec:QSOCatalog}

Because there are relatively few previously observed quasars in the desired BOSS redshift range ($z>2.2$) with sufficiently small flux errors to precisely describe the quasar color locus, for our purposes the QSO Catalog is generated by a Monte Carlo technique \citep{Hennawi2010} to provide a less biased and more complete sample than is available from the SDSS quasar catalog.  The Monte Carlo simulation uses a model of the quasar luminosity function based on the studies by \cite{Jiang2006} to compute the density of quasars as a function of redshift and $i-$magnitude.  The \cite{Jiang2006} luminosity function is used because it extends fainter than the luminosity function of \cite{Richards06} and thus better matches the high redshift quasars in the BOSS redshift regime.  SDSS Data Release 5 spectroscopically confirmed quasars \citep[DR5QSO;][]{SDSSDR5} are the photometric inputs to the Monte Carlo.  The simulation generates 9.94 million unique ($i$-magnitude, redshift) pairs down to $i =22.5$ ($0.5-$mag fainter than the BOSS magnitude limit) with a distribution given by the luminosity function.  Each simulated quasar ($O'$) is then matched to the SDSS quasar (DR5QSO) with the nearest redshift to $O'$.  The SDSS photometry of the DR5QSO quasar is rescaled such that its $i-$magnitude matches that of the simulated quasar ($O'$). We assume that quasar colors are not a function of magnitude in the redshift range of interest, and thus can be extrapolated in this manner to deeper fluxes.  Thus this technique preserves the relative fluxes while providing a more complete coverage of the flux space than only using known SDSS quasars.  Objects with redshifts in the range desired for BOSS targeting ($z > 2.2$) are included in the numerator of Eq.\;(\ref{probability}).  The location in $ugr$ color-color space of the $z>2.2$ objects in the QSO Catalog is shown by the blue contours in Fig.\;(\ref{fig:InputCatalogs}).

\subsection{Everything Else Catalog}
\label{subsec:EECatalog}

The Everything Else ($EE$) Catalog is generated using stacked SDSS ``Stripe 82'' imaging, allowing the construction of a large point source catalog with variability information and smaller errors than possible using single-epoch SDSS imaging. Stripe 82 is the $2.5^{\circ}$ wide region on the celestial equator between RA$=-45^{\circ}$ and RA$=45^{\circ}$ where SDSS repeatedly scanned.  Non-photometric data were ignored, and the photometric images were processed with a version of the SDSS photometric reduction pipeline similar to that in data release eight \citep{SDSSDR82011}.  The photometric depth is $r \sim 22.5$ magnitude ($5\sigma$) for point sources, with high completeness and accurate star-galaxy separation to $r \sim 22$ magnitude. These data were combined at the catalog level to produce co-added PSF photometry. Typically 20 observations were included for each object, resulting in a co-added catalog with typical errors of 6.1\%, 2.4\%, 3.0\%, 7.1\% and 27\% at 22nd magnitude in the $u, g, r, i, z$ filters, respectively.

\begin{figure}[t!]  \vspace{-0.2cm}
\begin{center}
      \includegraphics[height=6.5cm] {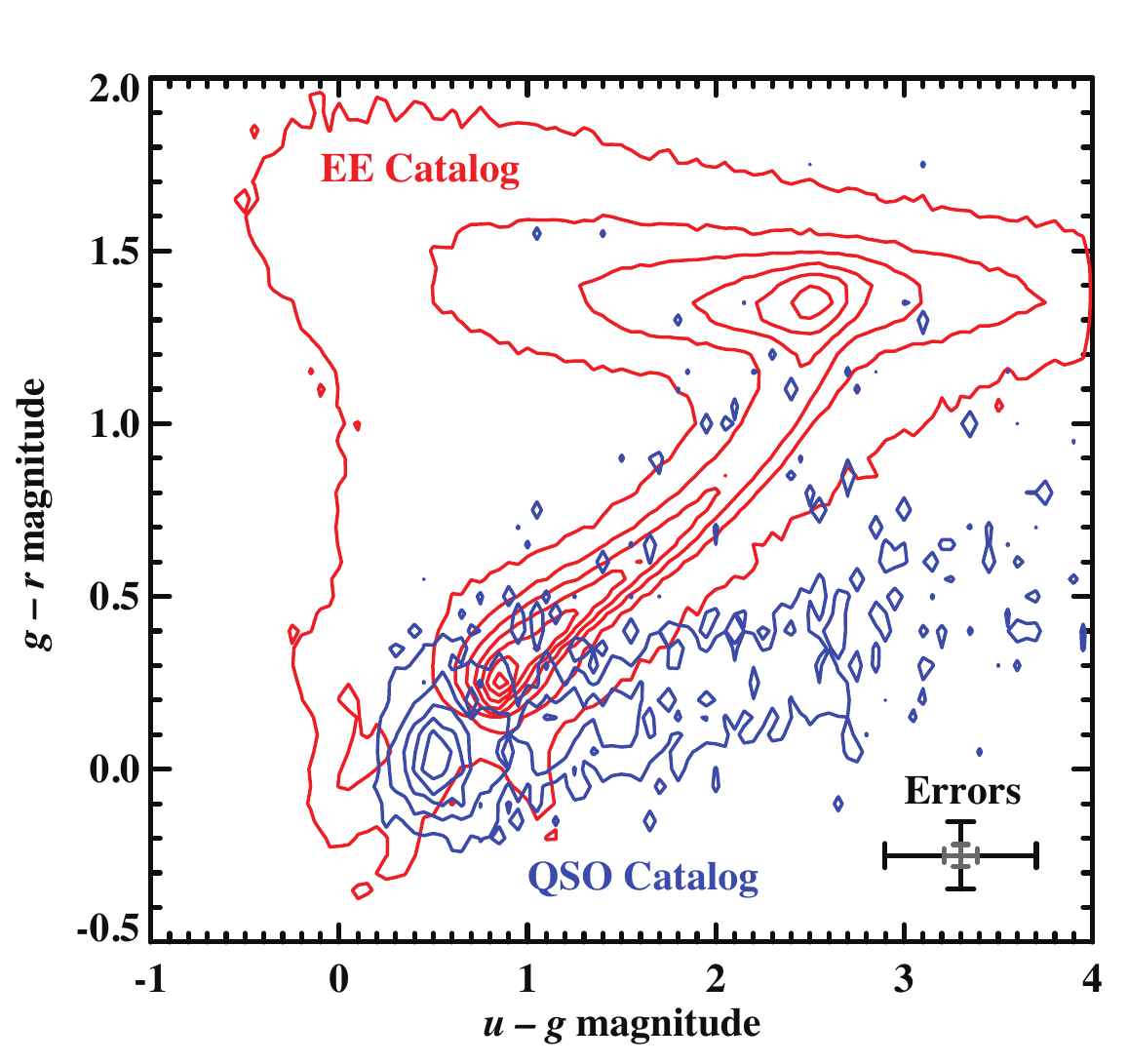}
     \caption[]
     {Contour plot of the $u-g$ and $g-r$ colors of the Everything Else (\emph{red}) and QSO (\emph{blue}) Catalogs.  The region of overlap, where target selection becomes difficult, is at $u-g \approx 1$ and $g-r \approx 0$.  The error bars are the SDSS single-epoch $g-r$ and $u-g$ magnitude errors at g=22 (\emph{black}) and g=20 (\emph{grey}).}  \vspace{-0.1cm}
\label{fig:InputCatalogs}
\end{center}
\end{figure}

The $EE$ Catalog is further trimmed to a clean sample of non-variable point sources for inclusion in the likelihood calculations of Eq.\;(\ref{probability}).  The 23.9\% of objects that are blended with neighboring objects are rejected, thus reducing the effective footprint of this catalog from 225 deg$^2$ to 171.2 deg$^2$.  Objects with high variability are explicitly excluded from the catalog under the presumption that these are dominated by quasars \citep{Schmidt10}, and we explicitly add quasars into the numerator and denominator of Eq.\;(\ref{probability}) such that the computed probability remains in the range [0,1].  These variable objects are identified as those with a reduced $\chi^2$ of the fit to a constant $r$-band flux exceeding 1.4.  This reduces the catalog to an effective area of 150 deg$^2$.  The result is a catalog with 1,042,262 photometric fluxes that represent all non-quasar types of objects.  We determined the contamination of $z > 2.2$ quasars in this set is less than 0.5\% by comparing this catalog with those for which we have spectra.  In Fig.\;(\ref{fig:InputCatalogs}) the red contours show the $urg$ color-color space of the objects in the $EE$ catalog.

\section{BOSS Data \& Likelihood Performance}
\label{sec:commissioning}

\subsection{BOSS Stripe 82 Data}
\label{subsec:commissioning}
In September of 2009, BOSS started taking spectroscopic data.  During the first year of data taking, several target selection methods were employed. In addition to likelihood method, three other selection techniques were deployed:  the KDE method developed to classify quasars by separating them from stars in color space \citep{Richards2004}, an ``extreme-deconvolution'' method (\cite{Bovy2011}, Section~\ref{subsec:XDQSO}) and a new approach based on artificial neural networks \citep{Yeche2009}.   Previously spectroscopically confirmed quasars, as well as objects with high variability \citep{PD11} over consecutive Stripe 82 runs were also targeted during this time.

Stripe 82 target selection used co-added catalog data from SDSS as the potential target fluxes.  Because the co-added photometry has a higher signal-to-noise ratio than any single-epoch data run and the target fiber density in this region was higher than the rest of the survey, BOSS QSO completeness is highest in this region.  Once observed, all of the quasar targets were automatically classified and then visually examined.  

Based on the objects selected in Stripe 82, we found that the performances of the four methods were not identical as a function of the magnitude and redshift of the objects \citep{Ross11}.  This behavior is likely due to the different strategies adopted in the training of the methods. 

\begin{figure}[b!]
\begin{center}
	\includegraphics[height=5.4cm] {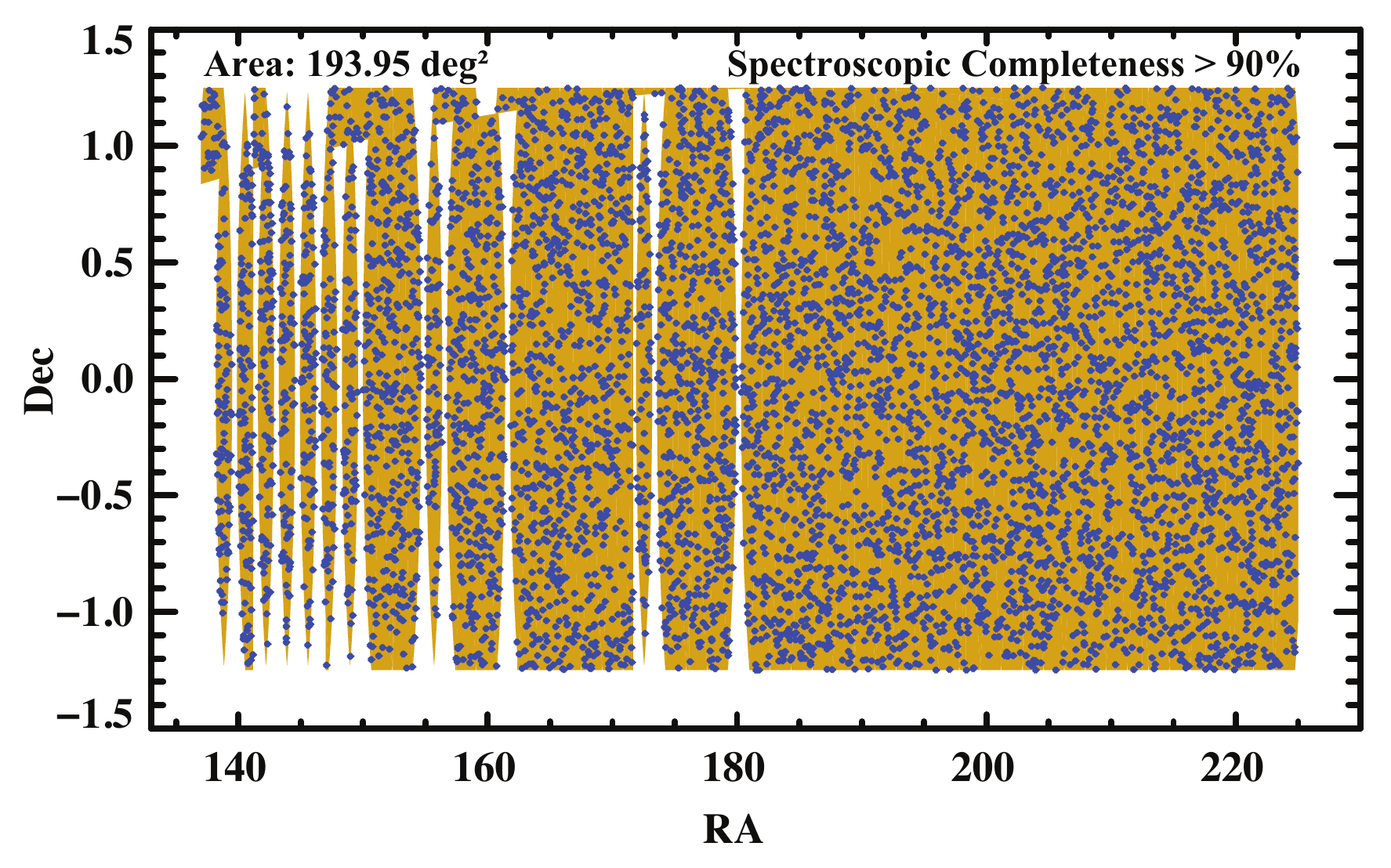}
	\caption[]
      {Right Ascension (RA) versus Declination (Dec) of BOSS QSO data used for the likelihood method testing and luminosity function testing.   Testing was done in the Stripe 82 calibration band with regions of high ($>90\%$) spectroscopic completeness.  The blue points are spectroscopically confirmed quasars and the yellow regions are the sky tiles that were observed.  Note that the vertical and horizontal scales are not the same.} \vspace{-0.5cm}
\label{fig:adamsfig}
\end{center}
\end{figure}

\subsection{Likelihood Performance}
\label{subsec:performance}
Although we targeted a number of tiles for spectroscopy during the first year of data taking, observational success was varied. Due to a combination of poor observing conditions and equipment glitches, spectroscopic completeness (the fraction of total spectroscopic observations in a tiling region which yielded a high confidence spectroscopic identification upon visual inspection) was a strong function of the region in which a target was tiled. In this paper, we only test our method using observations in Stripe 82 regions with a spectroscopic completeness of $> 90$\%. In Fig.\;(\ref{fig:adamsfig}) we show the tiles used for testing.

\begin{figure*}[t!]
\begin{center}
	\includegraphics[height=5.3cm] {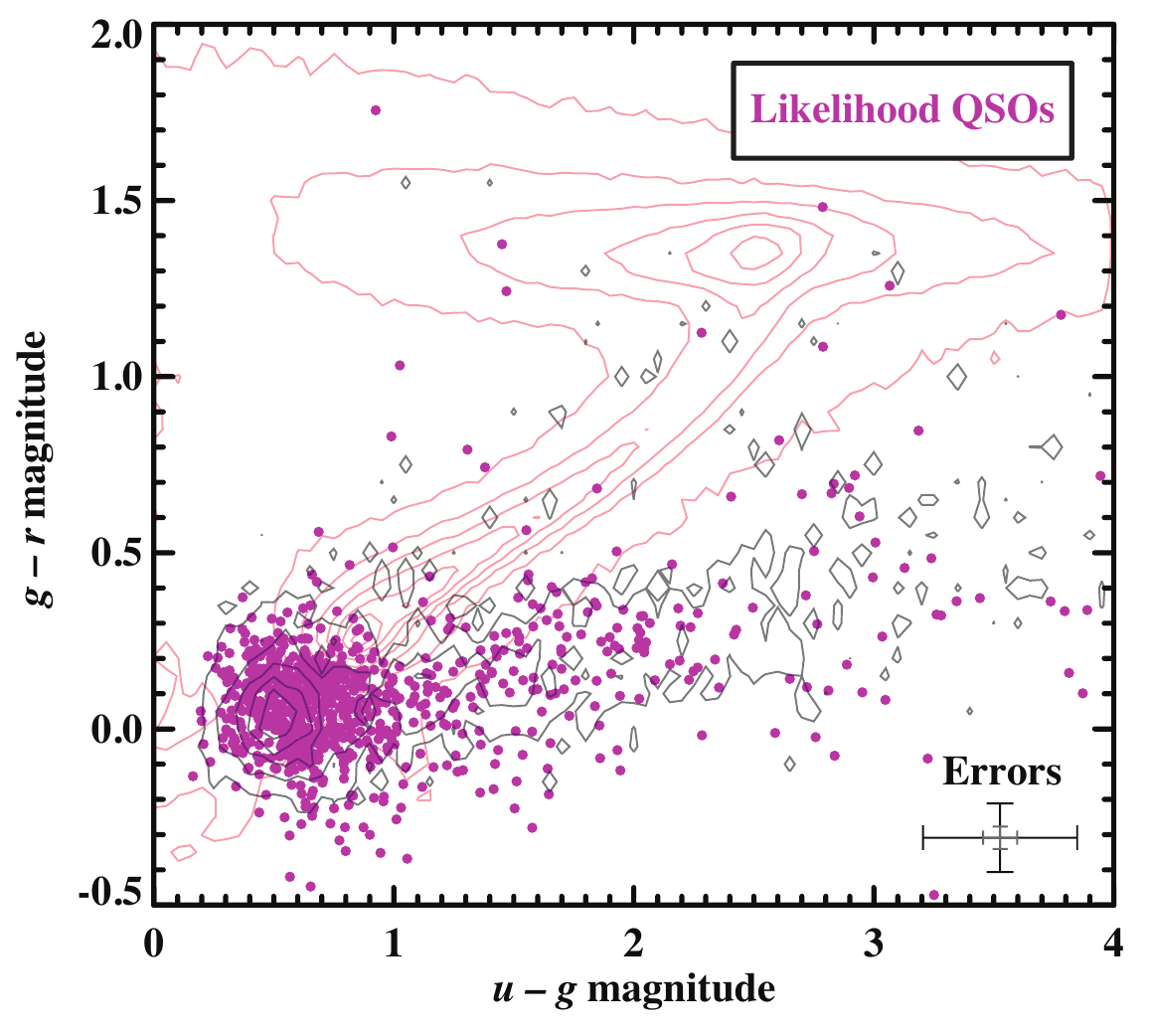}
	\includegraphics[height=5.3cm] {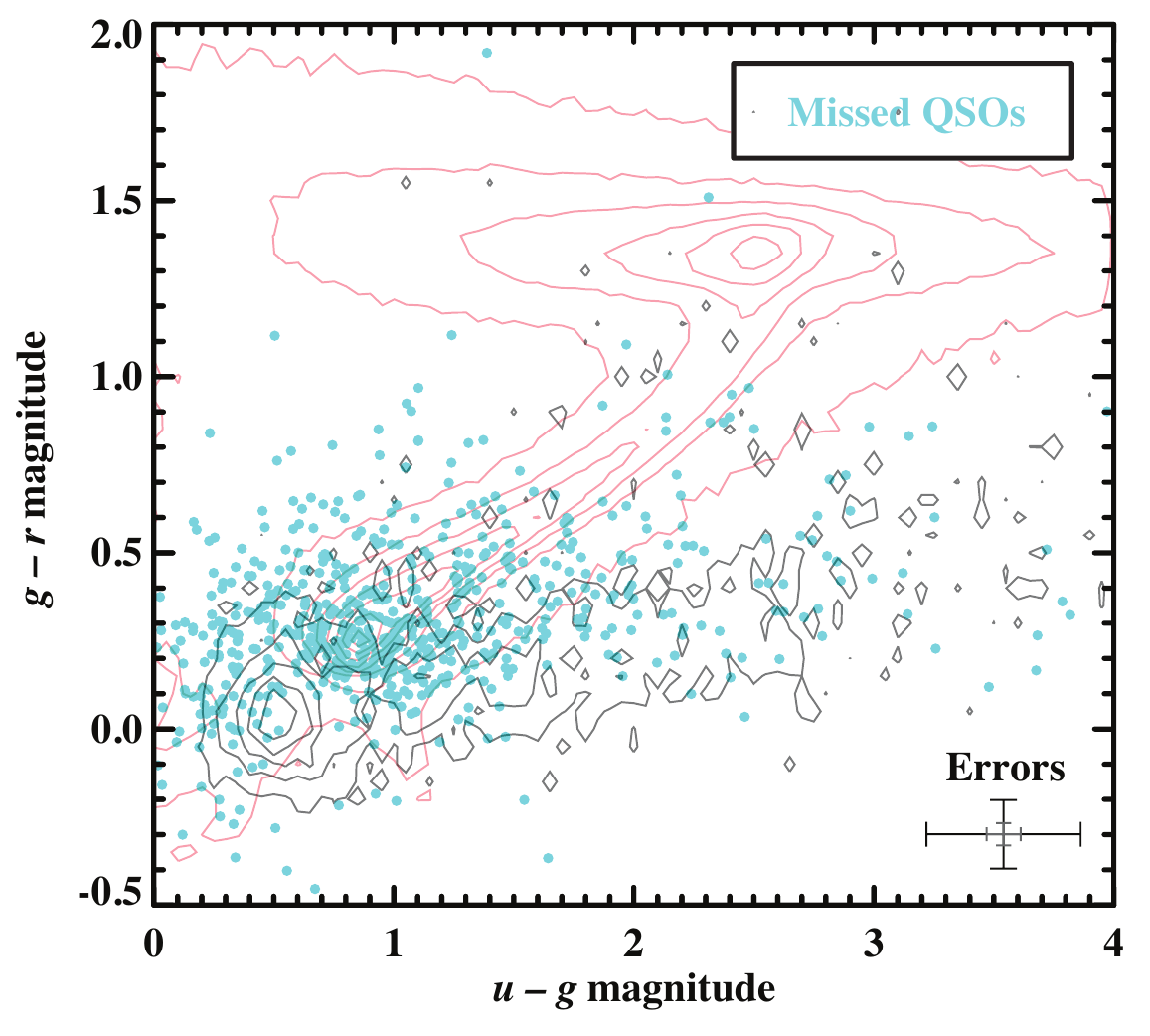}
	\includegraphics[height=5.3cm]{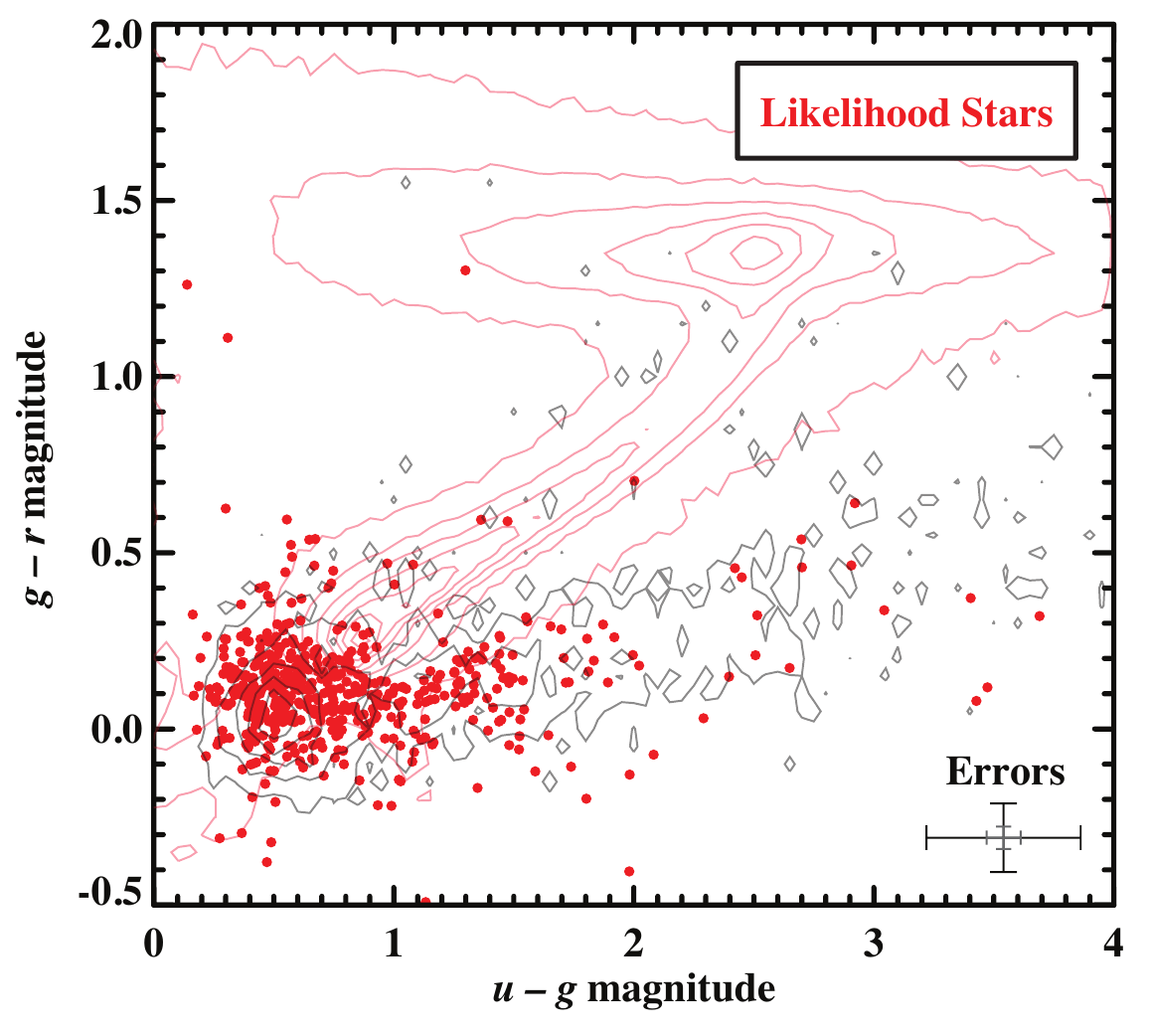}
	\caption[]
      {Color-color diagrams of BOSS QSOs recovered by the likelihood method (\emph{magenta}), false-negative QSOs that were not targeted (missed) by likelihood method (\emph{cyan}),  and false-positive stars that were wrongly targeted by likelihood method (\emph{red}).  These plots show recovered/missed ($z>2.2$) QSOs.  It is clear when comparing these plots with Fig.\;(\ref{fig:InputCatalogs}) that the problematic region for likelihood targeting is where the two catalogs overlap near $u-g=1$, $g-r=0.25$. For context the $QSO$ Catalog and $EE$ Catalog contours plot from Fig.\;(\ref{fig:InputCatalogs}) are included in the above plots.  The error bars are the SDSS single-epoch $g-r$ and $u-g$ magnitude errors at $g=22$ (\emph{black}) and $g=20$ (\emph{grey}).  The targeting decisions were computed in flux space rather than the color space shown in the figures.} \vspace{-0.3cm}
\label{fig:likelihoodQSOs}
\end{center}
\end{figure*}

\begin{figure}[b!] \vspace{0.3cm}
\begin{center} 
	\includegraphics[height=6.5cm]{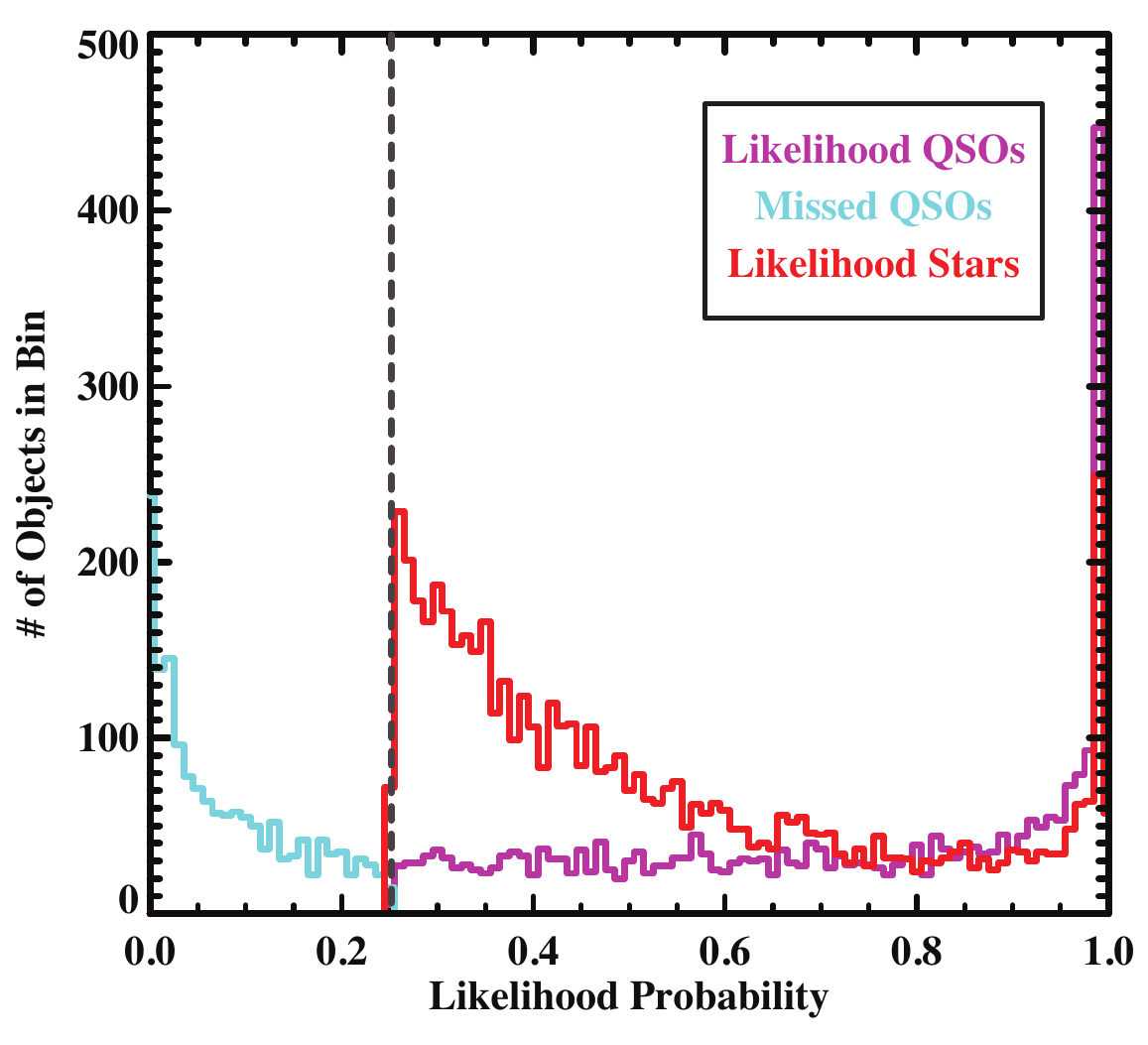}   
	\caption[]
      {The probability ($\mathcal{P}$) distributions of the likelihood method recovered QSOs (magenta, 4617 total), false-negative QSOs that were missed by the likelihood method (cyan, 1566 total), and false-positive stars that were incorrectly targeted by likelihood method (red, 5743 total).  The vertical gray dashed line shows the likelihood $\mathcal{P}$ threshold used for targeting ($\mathcal{P} > 0.245$).  The spike around $\mathcal{P} = 0$ in the cyan curve are quasars that fall in the midst of the stellar locus and therefore are found by the method to have a very low probability of being QSOs.  Most of these quasars are targeted because they are previously spectroscopically confirmed QSOs or by their flux variability.  The likelihood distribution of the probabilities for the untargeted stars (true-negative) are not included in the plot, but constitute an additional 742,662 objects.}\vspace{-0.3cm}
\label{fig:likelihoodprobabilitydistribution}
\end{center}
\end{figure}

To test the performance of our likelihood method, we calculated probabilities using Eq.\;(\ref{probability}) on single-epoch data in regions of Stripe 82 with high spectroscopic completeness and compared that target list with the BOSS ``truth table'' (which includes targets from all targeting methods, quasars targeted using variability, and all previously known quasars).  This is a fair test because targeting in this region was conducted using co-added photometry and thus we are not testing the likelihood method on a region that was targeted with the same photometry. 

Likelihood probabilities were calculated for 592,847 objects; of those, the top 7,757 likelihoods were selected $(\mathcal{P}>0.245)$ for a target density of 40 objects per deg$^2$.  Fig.\;(\ref{fig:likelihoodprobabilitydistribution}) shows the distribution of the probabilities for the recovered\footnote[2]{We define recovered/missed QSOs to be quasars in the desired BOSS redshift range ($z>2.2$).} and false-negative (missed) QSOs as well as for the false-positive stars (wrongly) targeted by the method.  We found an efficiency ($\mathcal{E}$ = Recovered QSOs / Total Targets) of 40\% and completeness ($\mathcal{C}$ = likelihood recovered QSOs / total BOSS recovered QSOs) of 65\%.  Fig.\;(\ref{fig:likelihoodQSOs}) shows $ugr$ color-color plots of BOSS quasars recovered (magenta) and missed (cyan) by the likelihood method as well as false positive contamination stars (red) that were targeted by the method. 

There is of course the inevitable trade-off between $\mathcal{E}$ and $\mathcal{C}$. The more fibers given to quasar targets, the more QSOs are found (greater $\mathcal{C}$), but the accuracy of targeting a quasar decreases (lower $\mathcal{E}$).  This is shown in Table\;(\ref{tab:commissioningresults}) where the rate of new targeted QSOs is shown to steadily decrease as a function of targets deg$^{-2}$.  

\begin{figure}
\begin{center}
	\includegraphics[height=6.5cm]{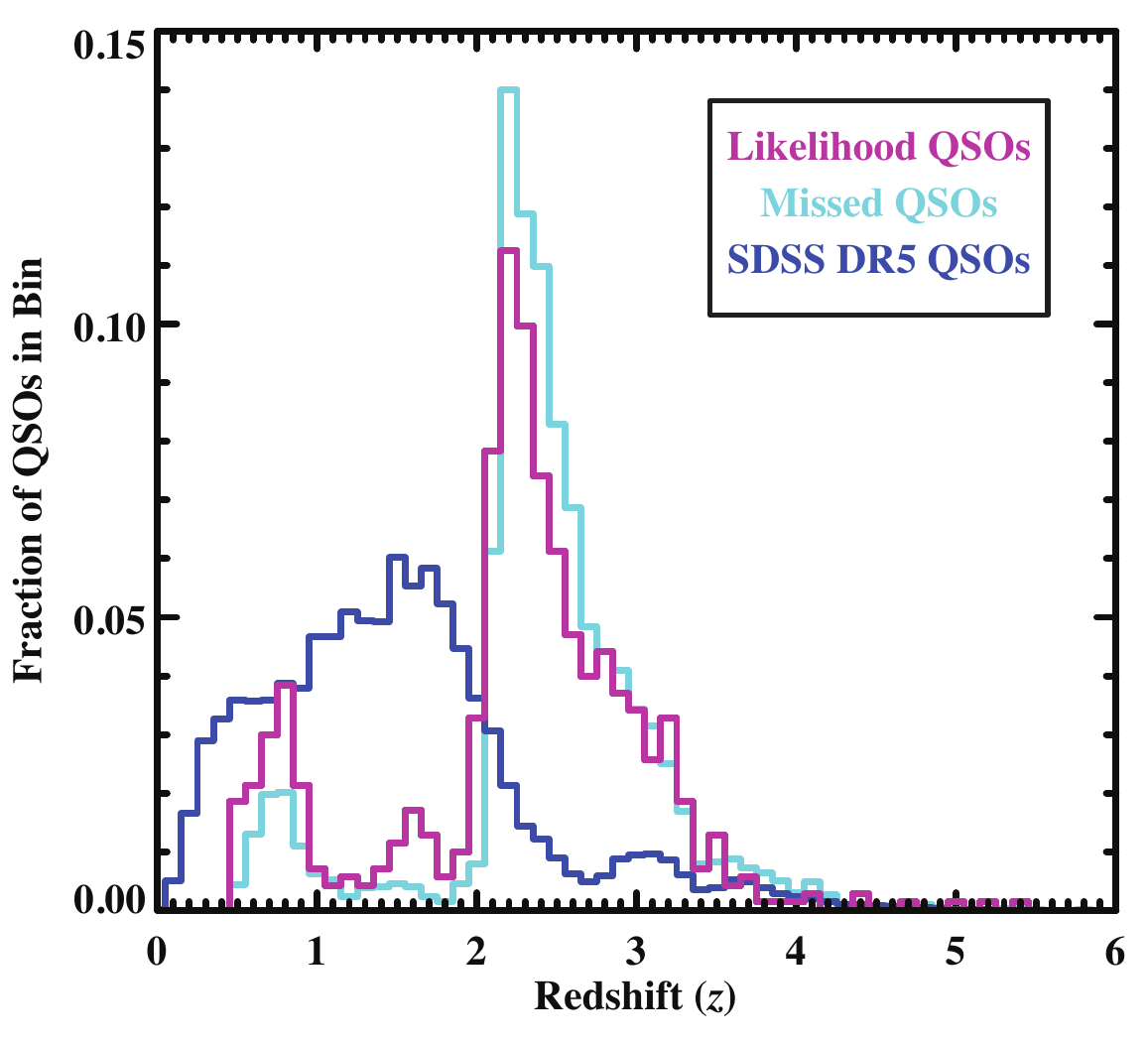}
	\caption[]
      {The redshift distributions of the likelihood method recovered QSOs (magenta) and false-negative QSOs that were not targeted (missed) by the likelihood method (cyan), compared with SDSS DR5 QSOs (blue).} \vspace{-0.3cm}
\label{fig:likelihoodredshiftdistribution}
\end{center}
\end{figure}

\begin{deluxetable*}{ccccccccccccccc}
\vspace{-0.8cm}
     \tablecaption{Likelihood Stripe 82 Results, \cite{Jiang2006} Luminosity Function
     \label{tab:commissioningresults}}
     \tablehead{\colhead{Targets} && \colhead{Likelihood $\mathcal{P}$}  && \colhead{Total}&& \colhead{QSOs} && \colhead{QSOs} &&& \colhead{$\mathcal{C}$} &&& \colhead{$\mathcal{E}$} \\
  \colhead{ per deg$^{2}$}  && \colhead{Threshold} && \colhead{Targets} && \colhead{Recovered} && \colhead{Missed} &&& \colhead{(\%)} &&& \colhead{(\%)}} 
    \tablecolumns{15}
     \startdata 
          $\,\;$5 &&  0.974 && $\,\;$969 && $\,\;$669 && 3811 &&& 15 &&& 69 \\
          10 && 0.833 && 1938 && 1276 && 3227 & && 28 &&& 66 \\
          20 && 0.535 && 3878 && 2166 && 2401 &&& 47 &&& 56 \\
          40 &&  0.245 && 7757 && 3087 && 1657 &&& 65 &&& 40 \\
          60 &&  0.136 && 11636$\,\;$ && 3595 && 1331 &&& 73 &&& 31 \\
          80 && 0.088 && 15515$\,\;$ && 3965 && 1108 &&& 78 &&& 26 \\
          100$\,\;$ && 0.063 && 19394$\,\;$ && 4219 && $\,\;$980 &&& 81 &&& 22 \\
          140$\,\;$ && 0.037 && 27152$\,\;$ && 4618 && $\,\;$806 &&& 85 &&& 17 
         \enddata
         \tablecomments{The $\mathcal{E}$ and $\mathcal{C}$ as a function of dedicated target fibers (targets deg$^{-2}$).  These values are for $z>2.2$ recovered/missed QSOs.  There is a trade-off: the more fibers given to targets, the more QSOs are found (greater $\mathcal{C}$), but the accuracy of finding a quasar decreases (lower $\mathcal{E}$). The values for threshold, $\mathcal{C}$ and $\mathcal{E}$ will of course depend on Galactic latitude \citep{Ross11}.  BOSS year-one data targeted using the likelihood method at 20 targets deg$^{-2}$ for the CORE sample.}
\end{deluxetable*}

By comparing the target objects to the catalog contours in Fig.\;(\ref{fig:likelihoodQSOs}), it is clear that the likelihood method fails mostly in the region of overlap between the two catalogs.   Fig.\;(\ref{fig:likelihoodredshiftdistribution}) shows the redshift distributions of the targeted and missed quasars and the limitation of SDSS DR5 catalog at $z>2$.  Table\;(\ref{tab:commissioningresults}) shows the detailed testing results.

\section{Testing, Improvements and Conclusions}
\label{sec:conclusions}

\subsection{Likelihood versus Color-Box}
\label{subsec:colorbox}

In order to see how our likelihood method performed against the traditional ``color-box'' selection, we compared the number of $z>2.2$ quasars the likelihood method was able to recover versus a simple color-box selection, using the BOSS data on Stripe 82 (the same data set as in Section~\ref{subsec:performance}). We note that our color-box, described below, is a relatively simple selection in only $(u-g)$ vs. $(g-r)$ magnitudes color-space, designed to adequately sample the location where $z\sim2.7$ QSOs reside. This color-box is not the same as the ``inclusion region'' from \cite{Richards02}(Section 3.5.2, Fig. 7) but the $(g-r) < 0.43\cdot(u-g)$ cut is inspired by their high-redshift color-selection.

The color cuts we used for our tests are:
\begin{equation}
\fontsize{9}{13}
\begin{aligned}[]
\label{ugcolorbox}
[((g-r) < -0.13 \cdot &(u-g) - A) \text{ and } \\
((u-g) > 0.3) &\text{ and } (r < 22.0)] \\
\text{ or }\\  
[((g-r) < 0.43 \cdot &(u-g) + B) \text{ and } \\
(0.3 < (u-g) < 2.&0) \text{ and } (r < 22.0)]
\end{aligned}
\end{equation}
where we offsets A and B in Eqn.\;(\ref{ugcolorbox}) are defined as:
\begin{equation}
\label{offsets}
\fontsize{9}{13}A= (0.01\cdot k) - 0.32, \;\;\;\;\;\;\;\;\;\; B= -(0.01\cdot k) - 0.28
\end{equation}
and $k$ is an integer varying from 0 to 29, $k = [0,1,2, \cdots, 27,28,29]$.

\begin{figure}
\begin{center}
	\includegraphics[height=6.5cm]{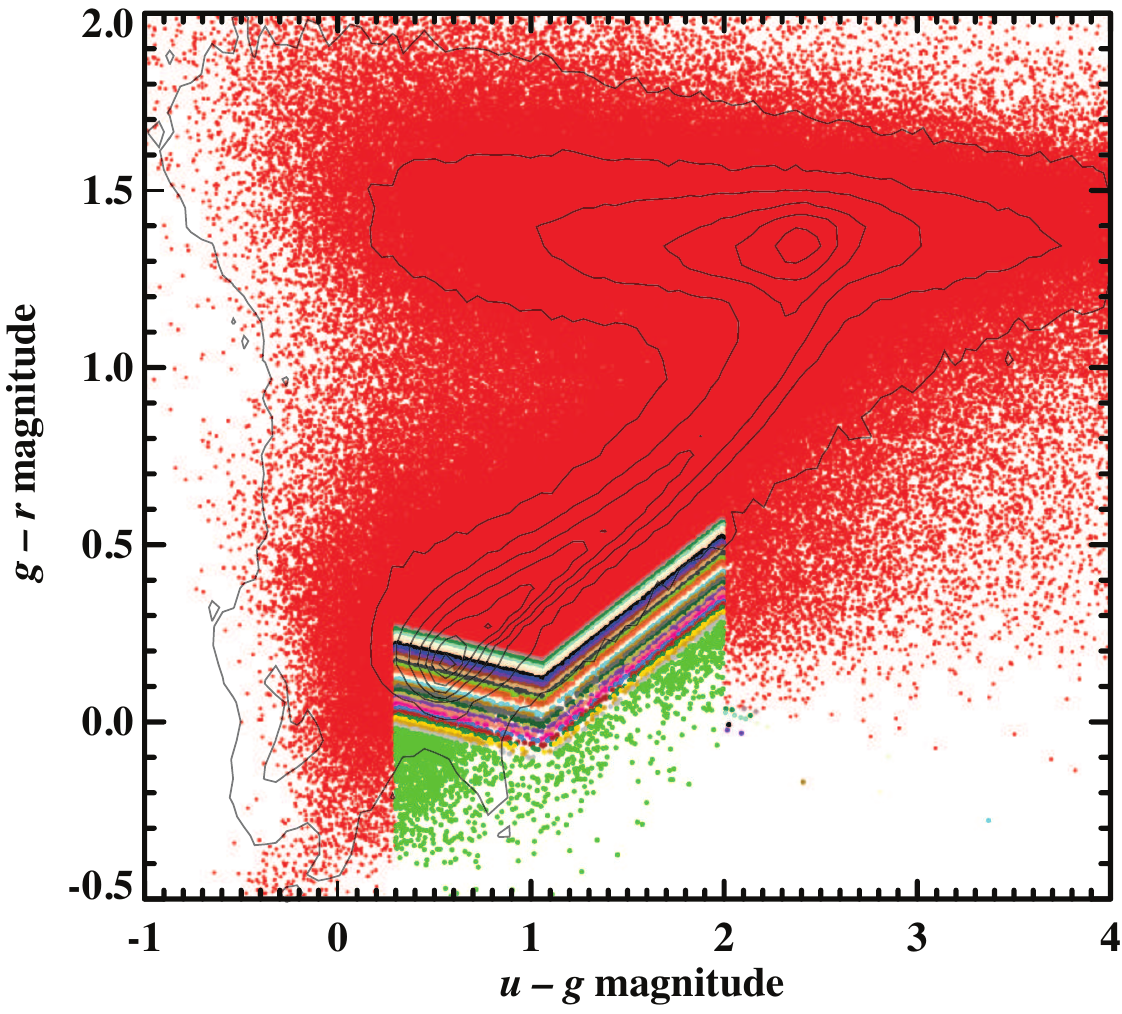}
	\caption[]
      {A color-color diagram showing the cuts applied for color-box selection.  The red points are potential targets that were not selected by the color-box method.  The other colored regions are targets selected by the cuts in Eqn.\;(\ref{ugcolorbox}) with the different offsets from Eqn.\;(\ref{offsets}).  The color-box selection cuts with the most targets is at the top (closest to the stellar locus, shown as grey contours), and the cuts get more restrictive moving towards the x-axis.}
\label{fig:colorcolorbox}
\end{center}
\end{figure}

In Figure\;(\ref{fig:colorcolorbox}) we show the above color cuts applied to the Stripe-82 potential quasar targets.  The red points are targets that are not selected by the color-box.  The points that are other colors at the bottom center of the figure are targets that were selected by the cuts in Eqn.\;(\ref{ugcolorbox}) with the different offsets from Eqn.\;(\ref{offsets}).  The color-box selection cuts with the most targets ($k=0$) are closest to the stellar locus (shown as grey contours).  The cuts then get more restrictive going down towards the x-axis, where less and less targets are selected.

\begin{figure}
\begin{center}\vspace{-0.7cm}
	\includegraphics[height=6.5cm]{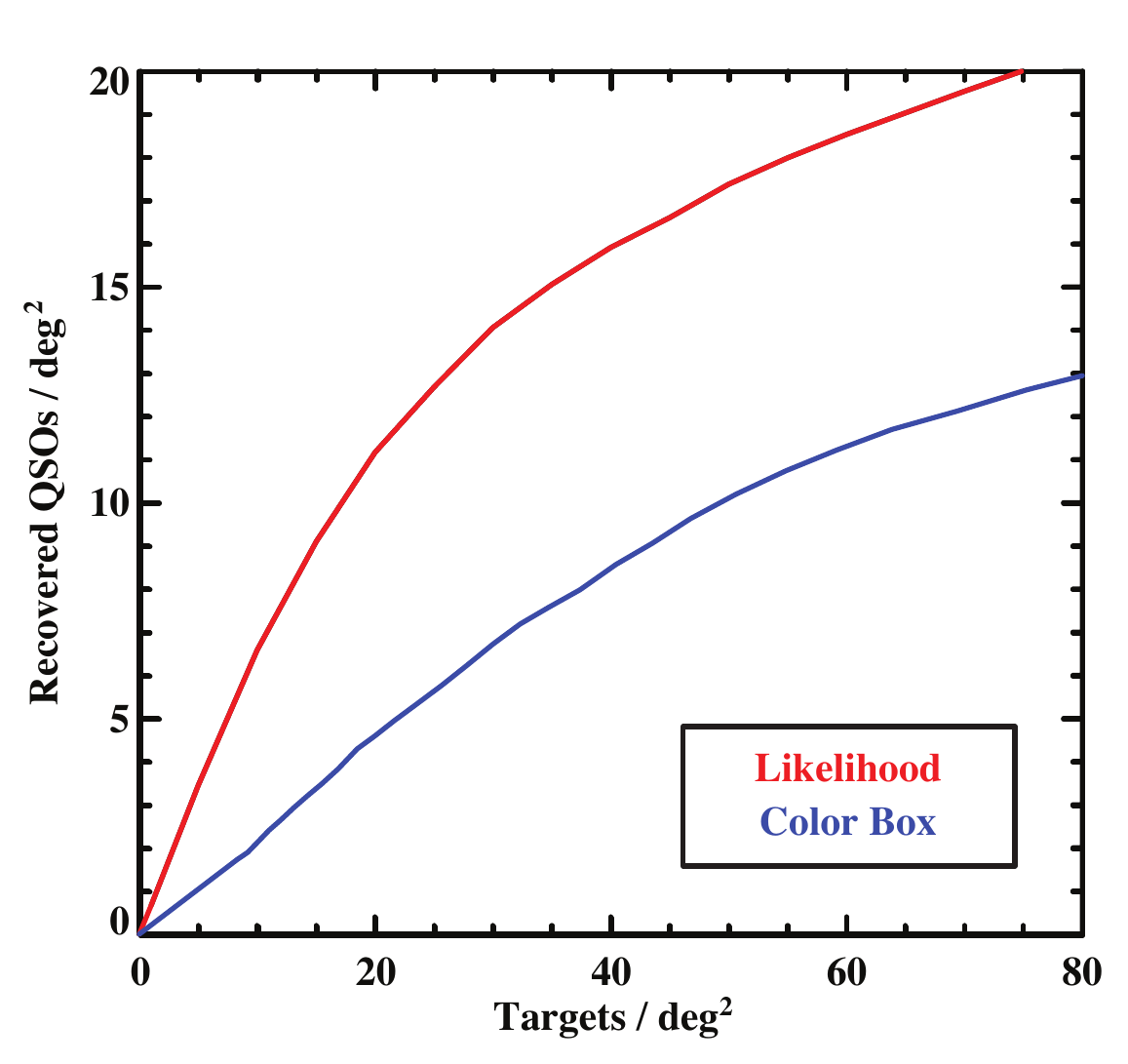}
	\caption[]
      {The number of BOSS $z>2.2$ quasars recovered as a function of targets deg$^{-2}$ for the likelihood method (red) and a traditional color-box technique (blue).  Likelihood out performs the color-box selection method by recovering over twice as many BOSS quasars at 20 targets deg$^{-2}$.}\vspace{-0.5cm}
\label{fig:likecolorbox}
\end{center}
\end{figure}
 
\begin{figure*}[t!]\vspace{-0.2cm}
\begin{center}
    \includegraphics[height=5.4cm]
     {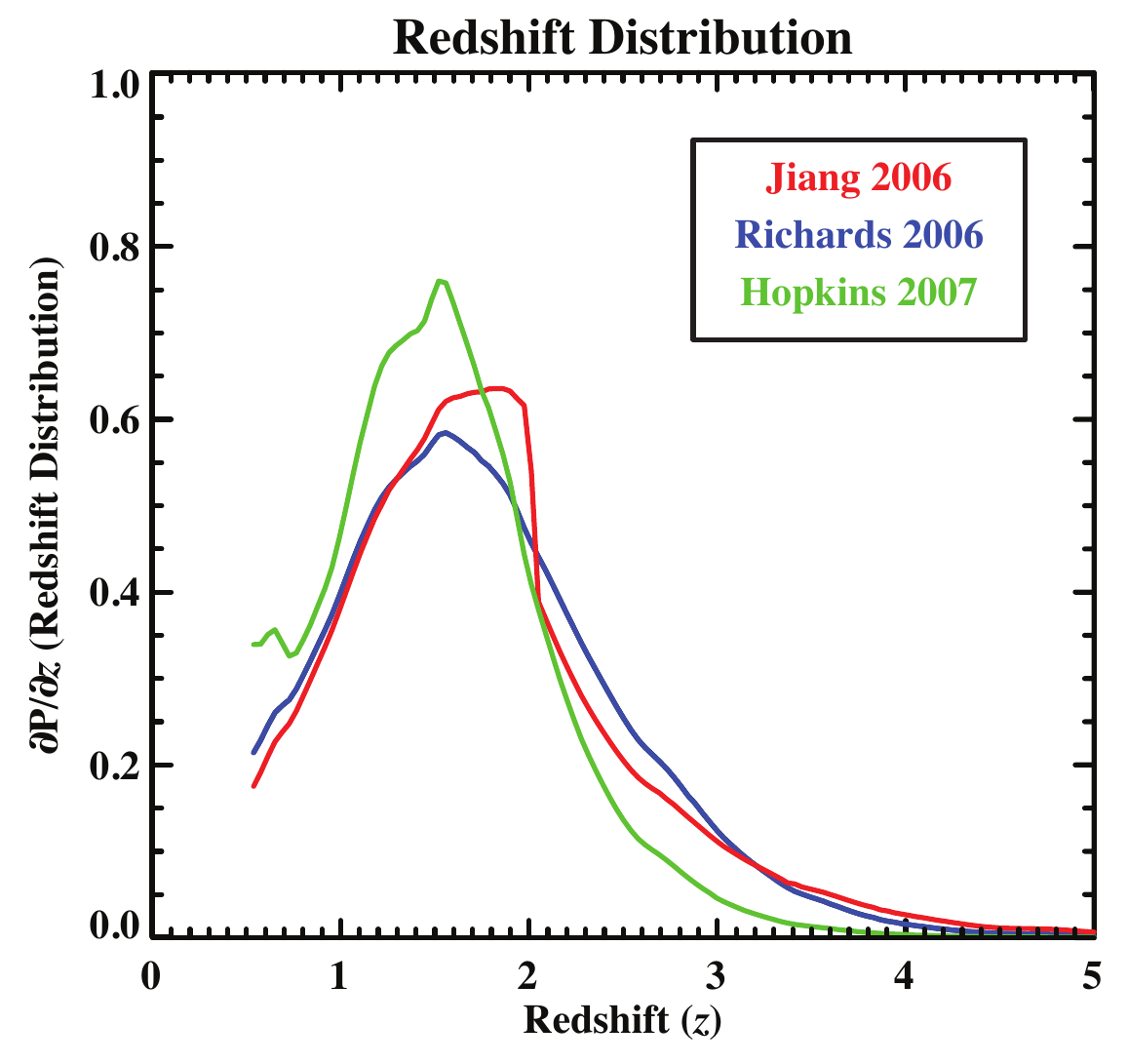}
    \includegraphics[height=5.4cm]
     {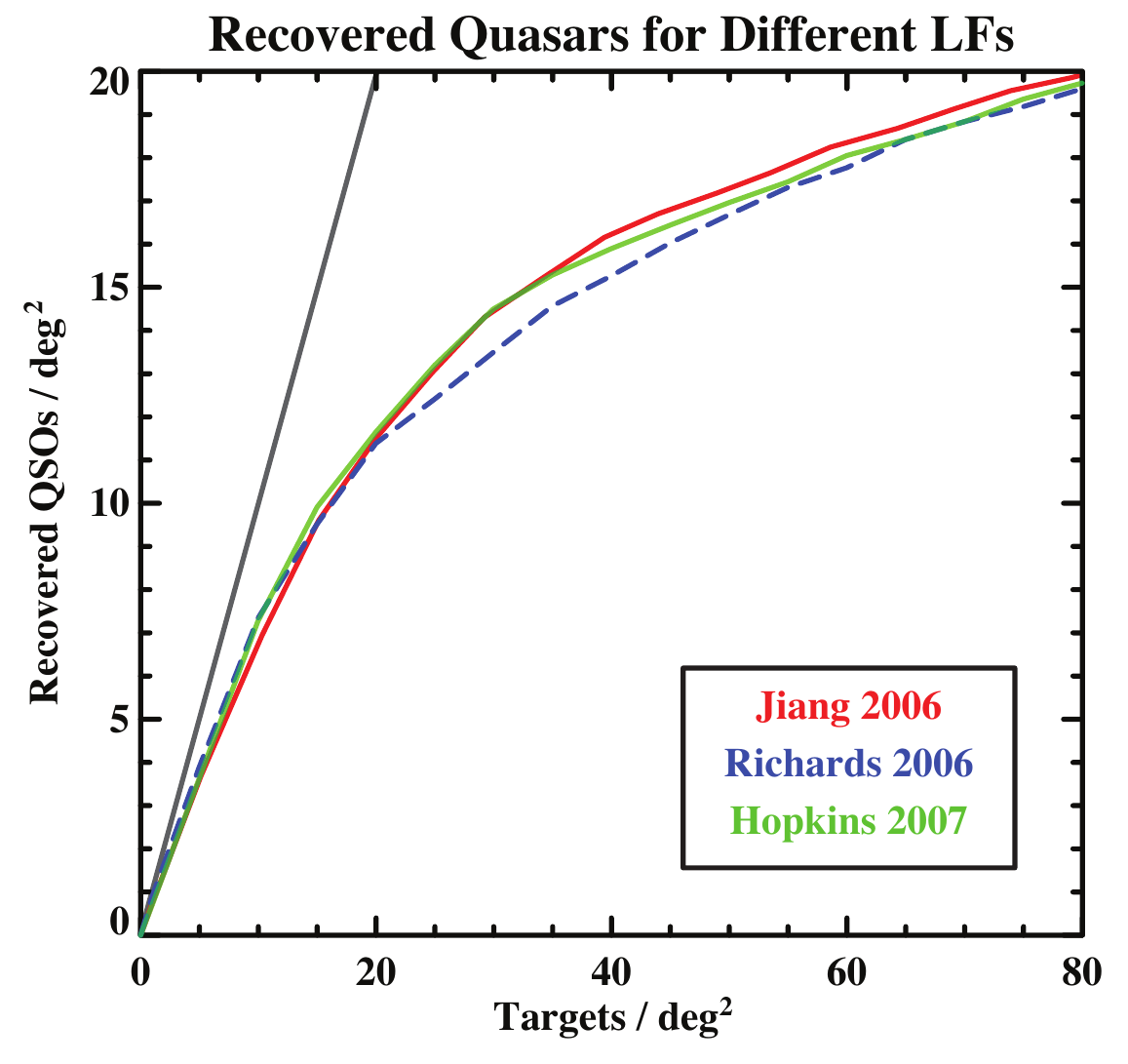}
     \includegraphics[height=5.4cm]
     {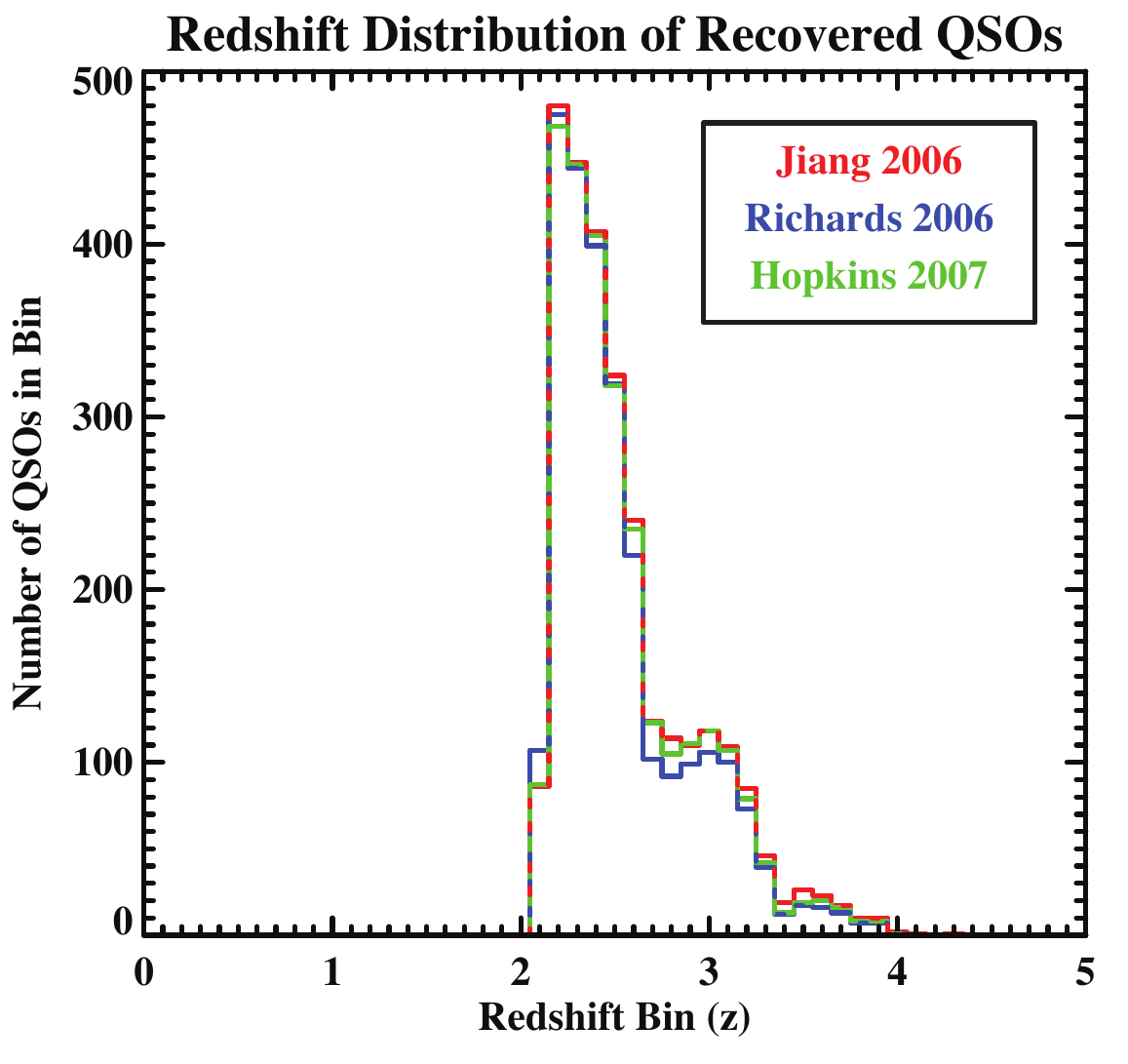}
     \caption[]
      {(\emph{Left}) - Redshift distribution for \cite{Jiang2006} (red), \cite{Richards06} (blue) and \cite{HRH07} (green)  luminosity functions.  (\emph{Center}) - The number of BOSS $z>2.2$ quasars recovered as a function of targets deg$^{-2}$ for the three luminosity functions (i.e. different priors). The performance of all three LFs is almost identical for target densities up to $\sim20$ targets deg$^{-2}$, at which point the Richards model starts to perform sightly worse.  The gray line shows 100$\%$ efficiency, and emphasizes that very high efficiency is achieved if a small number of targets are selected.  (\emph{Right}) - The redshift distributions of the recovered $ z>2.2$ quasars for the three luminosity functions.}\vspace{-0.5cm}
\label{fig:lumfunct}
\end{center}
\end{figure*}

 \begin{deluxetable*}{ccccccccccccccccc}\vspace{-0.9cm}
     \tablecaption{Likelihood Luminosity Function Testing
     \label{tab:lftestingresults}}
     \tablehead{
     \colhead{Luminosity} && \colhead{Targets} && \colhead{Likelihood $\mathcal{P}$}  && \colhead{Total}&& \colhead{QSOs} && \colhead{QSOs} &&& \colhead{$\mathcal{C}$} &&& \colhead{$\mathcal{E}$} \\
     \colhead{Function} &&  \colhead{per deg$^{2}$}  && \colhead{Threshold} && \colhead{Targets} && \colhead{Recovered} && \colhead{Missed} &&& \colhead{(\%)} &&& \colhead{(\%)}
     }
    \tablecolumns{17}
     \startdata
	Jiang et al. && 20 && 0.535 && 3878 && 2166 && 2401 &&& 47 &&& 56 \\
	&& 40 &&  0.245 && 7757 && 3087 && 1657 &&& 65 &&& 40 \\ [0.1cm]
	Richards et al. &&  20 && 0.237 && 3978 && 2089 && 2466 &&& 46 &&& 54 \\
	&&  40 && 0.079 && 7757 && 2939 && 1808 &&& 62 &&& 37 \\ [0.1cm]
	Hopkins et al. &&  20 && 0.383 && 3878 && 2154 && 2401 &&& 47 &&& 56 \\
	&&  40 && 0.158 && 7737 && 3046 && 1676 &&& 64 &&& 39 
     \enddata
         \tablecomments{Shows the completeness ($\mathcal{C}$) and efficiency ($\mathcal{E}$) as a function of dedicated target fibers (targets deg$^{-2}$) for three different luminosity functions. The three LFs tested are from \cite{Jiang2006}, \cite{Richards06}, and \cite{HRH07}. These values are for $z>2.2$ recovered/missed QSOs.}                  
   \end{deluxetable*}

Figure\;(\ref{fig:likecolorbox}) show the results of $z>2.2$ quasars recovered by the above color cuts compared to those recovered by the likelihood method.  This is consistent with the results in Table 8 of  \cite{Ross11DR9} which shows 6.45 mid-$z$ quasars are recovered from 20 targets deg$^{-2}$ using their (slightly different) color-box selection. Likelihood out performs the color-box selection method by recovering over twice as many BOSS quasars at 20 targets deg$^{-2}$.

\subsection{Luminosity Function Testing}
\label{subsec:lf}

We tested the performance of three different quasar luminosity functions (LFs) as inputs to the $QSO$ Catalog.  The LFs enter into the generation of the $QSO$ Catalog by determining the density of quasars as a function of redshift and $i-$band magnitude.  All the other details of the Monte Carlo remain the same as described in Section\;(\ref{subsec:QSOCatalog}).  The $EE$ Catalog is not dependent on these LFs so this catalog stays the same for these tests.  The three functions tested are from \cite{Jiang2006}, \cite{Richards06}, and \cite{HRH07}.   The inputs and results from this testing is shown in Fig.\;(\ref{fig:lumfunct}).  

The quasar redshift distributions for these three luminosity functions are shown in the left panel of Fig.\;(\ref{fig:lumfunct}).  The performance of the method did not change significantly for the three different LF priors.  Fig.\;(\ref{fig:lumfunct}, \emph{Center}) shows the number of quasars successfully recovered as a function of the number of dedicated QSO target fibers per deg$^2$.  Notice the shape of this function, the rate of newly recovered quasars drops off significantly beyond 40 targets deg$^{-2}$. The performance of all three LFs is essentially identical up to 20 targets deg$^{-2}$.  The redshift distributions of the recovered quasars are slightly different as shown in Fig.\;(\ref{fig:lumfunct}, \emph{Right}), so using different LF could be used to help tune the targeting redshifts.  This method applied to BOSS targeting is not sensitive to uncertainties in the luminosity function.

Ultimately it was decided that \cite{Jiang2006} was the best luminosity function for our purposes because it was more efficient at recovering $z>2.2$ QSOs than \cite{Richards06} and \cite{HRH07}.  More detailed values for the performance of these different luminosity functions can be seen in Table\;(\ref{tab:lftestingresults}).

While we are not currently making a proper comparison of the redshift distributions of BOSS quasars and the redshift distributions of these luminosity functions, a future improvement would be to use a luminosity function generated from the redshift distribution of BOSS quasars, properly adjusted for the targeting selection function imprinted upon it, as the input to the Monte Carlo to see if this approach improves target selection.  Another promising modification would be to add the photometry from BOSS quasars to the inputs to the Monte Carlo simulation.

\begin{figure*}[t!] \vspace{-0.2cm}
\begin{center}
    \includegraphics[height=5.4cm]
     {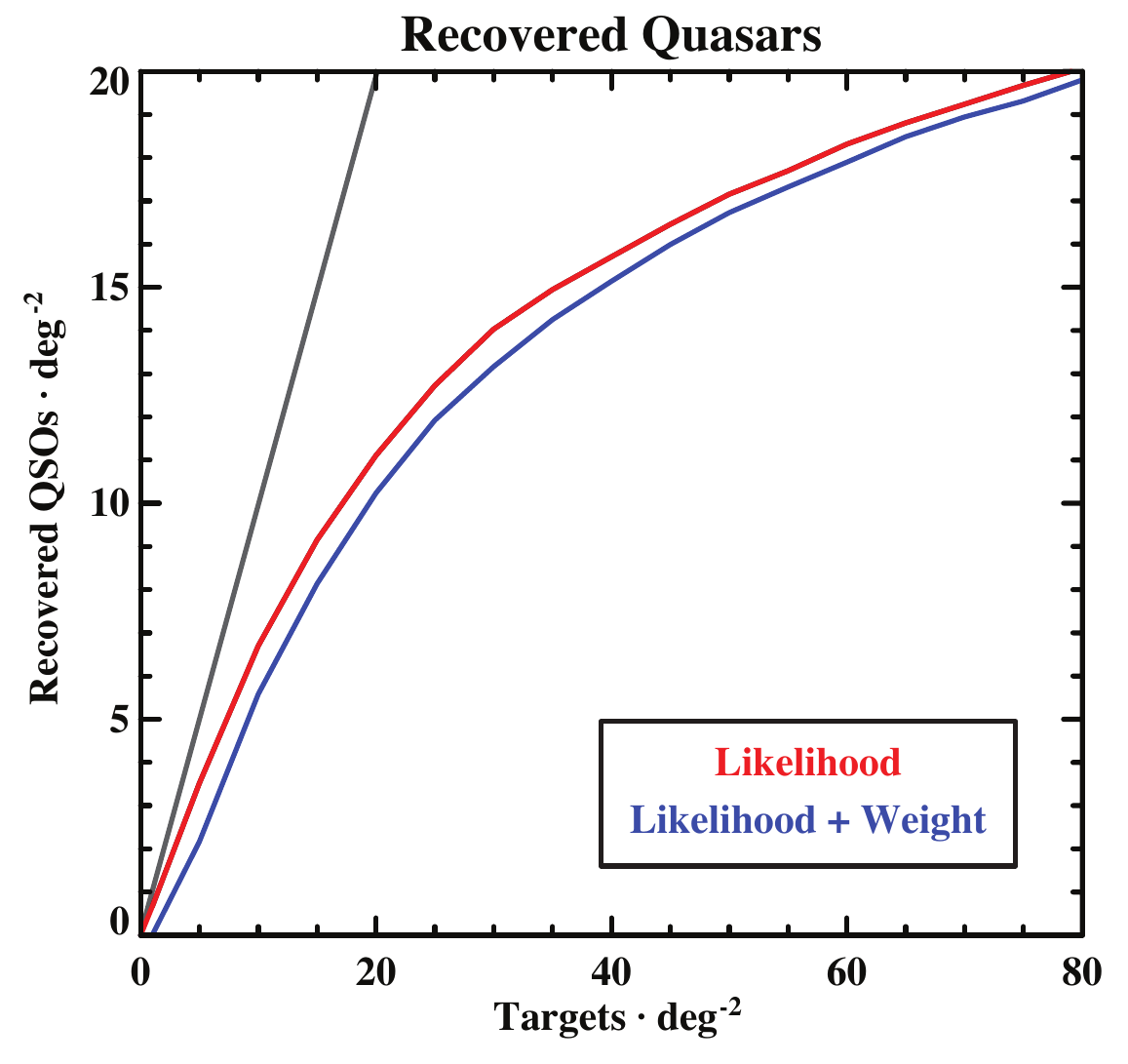}
    \includegraphics[height=5.4cm]
     {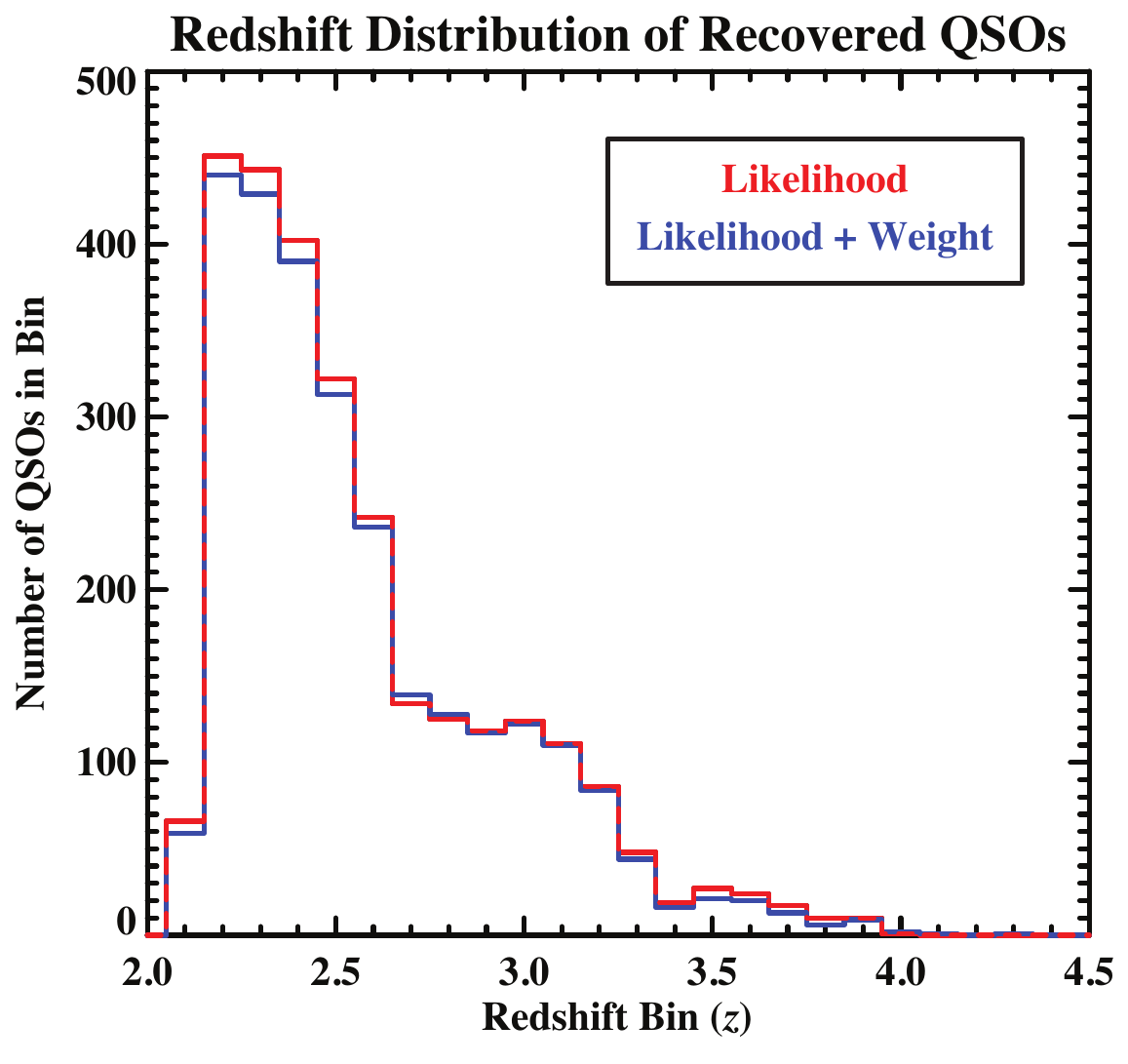}
     \includegraphics[height=5.4cm]
     {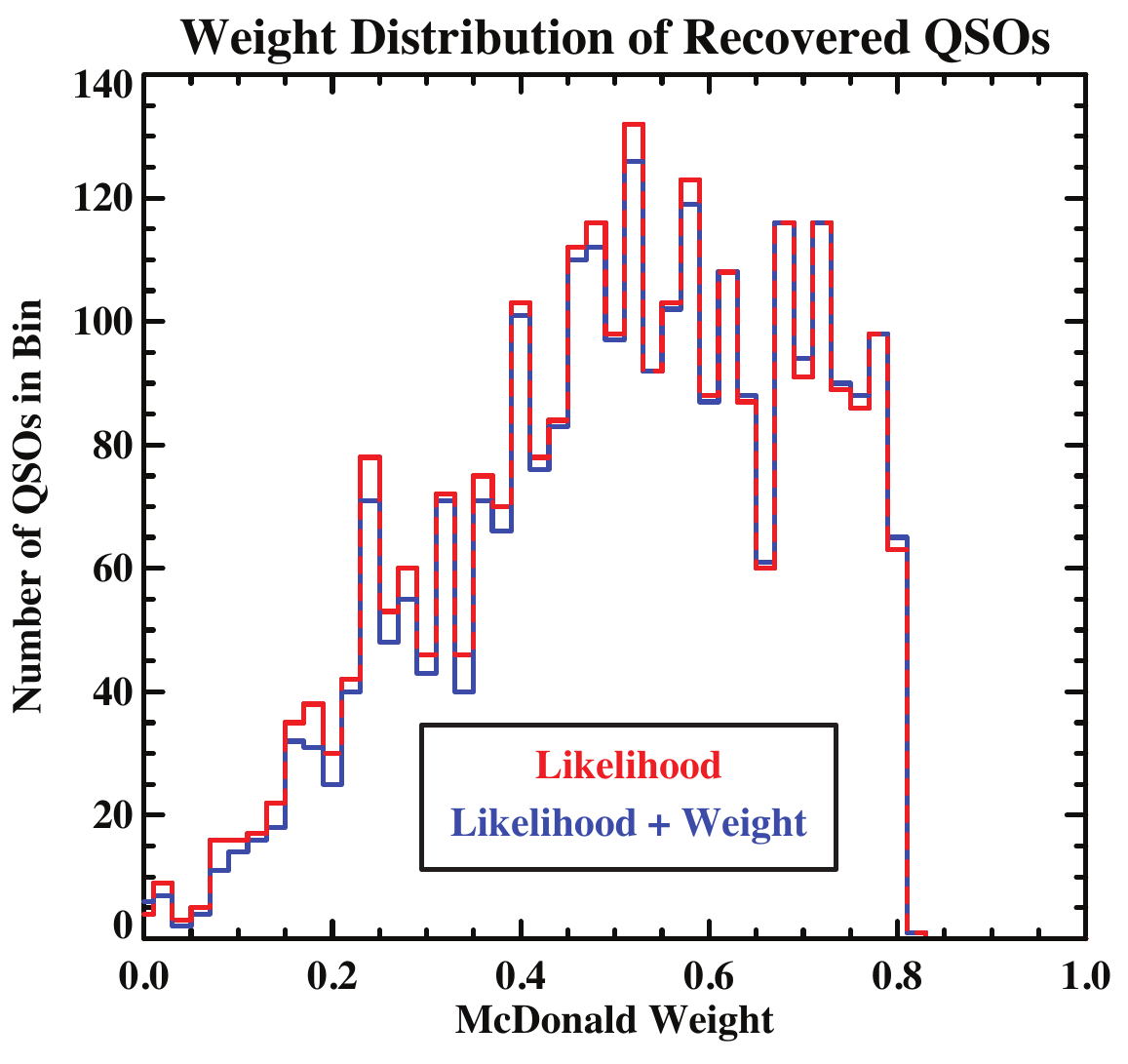}
     \caption[]
      {(\emph{Left}) - The number of BOSS $z>2.2$ quasars recovered as a function of targets deg$^{-2}$ for the likelihood method with and without weights.  The gray line shows 100$\%$ efficiency, and emphasizes that very high efficiency is achieved if a small number of targets are selected.  (\emph{Center}) - The redshift distributions of the recovered $ z>2.2$ quasars.  (\emph{Right}) - The weight distributions of the recovered $ z>2.2$ quasars.  Notice that using the likelihood + weights recovers QSOs with a higher BAO value.}
\label{fig:weightingFigs}
\end{center}
\end{figure*}

\begin{deluxetable*}{crrcccccccccc}\vspace{-1.2cm}
 \tabletypesize{\footnotesize}
     \tablecaption{McDonald Weights 
     \label{tab:weights}}
     \tablehead{\colhead{$r$-mag} & \colhead{} & \colhead{$z$ = \bf{2.0}$\:$} & \colhead{\bf{2.25}} & \colhead{\bf{2.5}} & \colhead{\bf{2.75}} & \colhead{\bf{3.0}} & \colhead{\bf{3.25}} & \colhead{\bf{3.5}} & \colhead{\bf{3.75}} & \colhead{\bf{4.0}} & \colhead{\bf{4.25}} & \colhead{\bf{4.5}}}
    \tablecolumns{13}
     \startdata 
\bf{17.5} && 0.00 & 0.487 & 0.822 & 0.713 & 0.570 & 0.441 & 0.334 & 0.250 & 0.162 & 0.162 & 0.162 \\ 
\bf{18.1} && 0.00 & 0.476 & 0.818 & 0.712 & 0.569 & 0.441 & 0.334 & 0.250 & 0.161 & 0.161 & 0.161 \\ 
\bf{18.7} && 0.00 & 0.464 & 0.814 & 0.710 & 0.568 & 0.440 & 0.333 & 0.250 & 0.161 & 0.161 & 0.161 \\  
\bf{19.3} && 0.00 & 0.446 & 0.807 & 0.708 & 0.567 & 0.439 & 0.333 & 0.250 & 0.161 & 0.161 & 0.161 \\  
\bf{19.9} && 0.00 & 0.411 & 0.794 & 0.704 & 0.564 & 0.438 & 0.332 & 0.249 & 0.161 & 0.161 & 0.161 \\ 
\bf{20.5} && 0.00 & 0.350 & 0.758 & 0.692 & 0.557 & 0.434 & 0.330 & 0.247 & 0.159 & 0.159 & 0.159 \\  
\bf{21.1} && 0.00 & 0.288 & 0.698 & 0.665 & 0.541 & 0.424 & 0.323 & 0.242 & 0.155 & 0.155 & 0.155 \\  
\bf{21.7} && 0.00 & 0.198 & 0.565 & 0.597 & 0.499 & 0.401 & 0.307 & 0.227 & 0.144 & 0.144 & 0.144 \\  
\bf{22.3} && 0.00 & 0.113 & 0.381 & 0.446 & 0.388 & 0.331 & 0.269 & 0.196 & 0.118 & 0.118 & 0.118 \\ 
\bf{22.9} && 0.00 & 0.041 & 0.187 & 0.254 & 0.247 & 0.231 & 0.173 & 0.109 & 0.061 & 0.061 & 0.061\\ 
\bf{23.5} && 0.00 & 0.008 & 0.045 & 0.054 & 0.049 & 0.046 & 0.026 & 0.015 & 0.004 & 0.004 & 0.004
     \enddata
         \tablecomments{Shown is a subsample of the values for the weights in Eqn.\;(\ref{likeQSOWeight}).  For a full length, downloadable table of weights, please see the electronic version of this paper.}                 
   \end{deluxetable*}\vspace{0.4cm} 

\vspace{-0.4cm}
\subsection{Weighted Likelihoods}
\label{subsec:weight}

We also tested adjusting Eq.\;(\ref{likeQSO}) to incorporate a weighting factor to optimize (in redshift-magnitude space) selection of objects with a high dark energy figure of merit \citep{DETF2006}.  This weighting is done by simply adding a factor ($w_{O'}$) inside the product based on the value of the QSO catalog quasar flux ($f^{O'}$) and redshift ($z$):
\begin{equation}
\label{likeQSOWeight}
\fontsize{9.0}{13}\mathcal{L}_{QSO}(\Delta z) =\!\!\!\!\!\!\!\! \displaystyle\sum_{O' \in QSO(\Delta z)} \displaystyle\prod_{f=\mathbf{f}} \frac{w_{O'}}{\sqrt{2\pi\sigma_{f}^2}} \exp{\left[-\frac{(f - f^{O'})^2}{2\sigma_{f}^2} \right]} 
\end{equation}

\begin{figure}[b!]
\begin{center}
	\includegraphics[height=6.5cm]{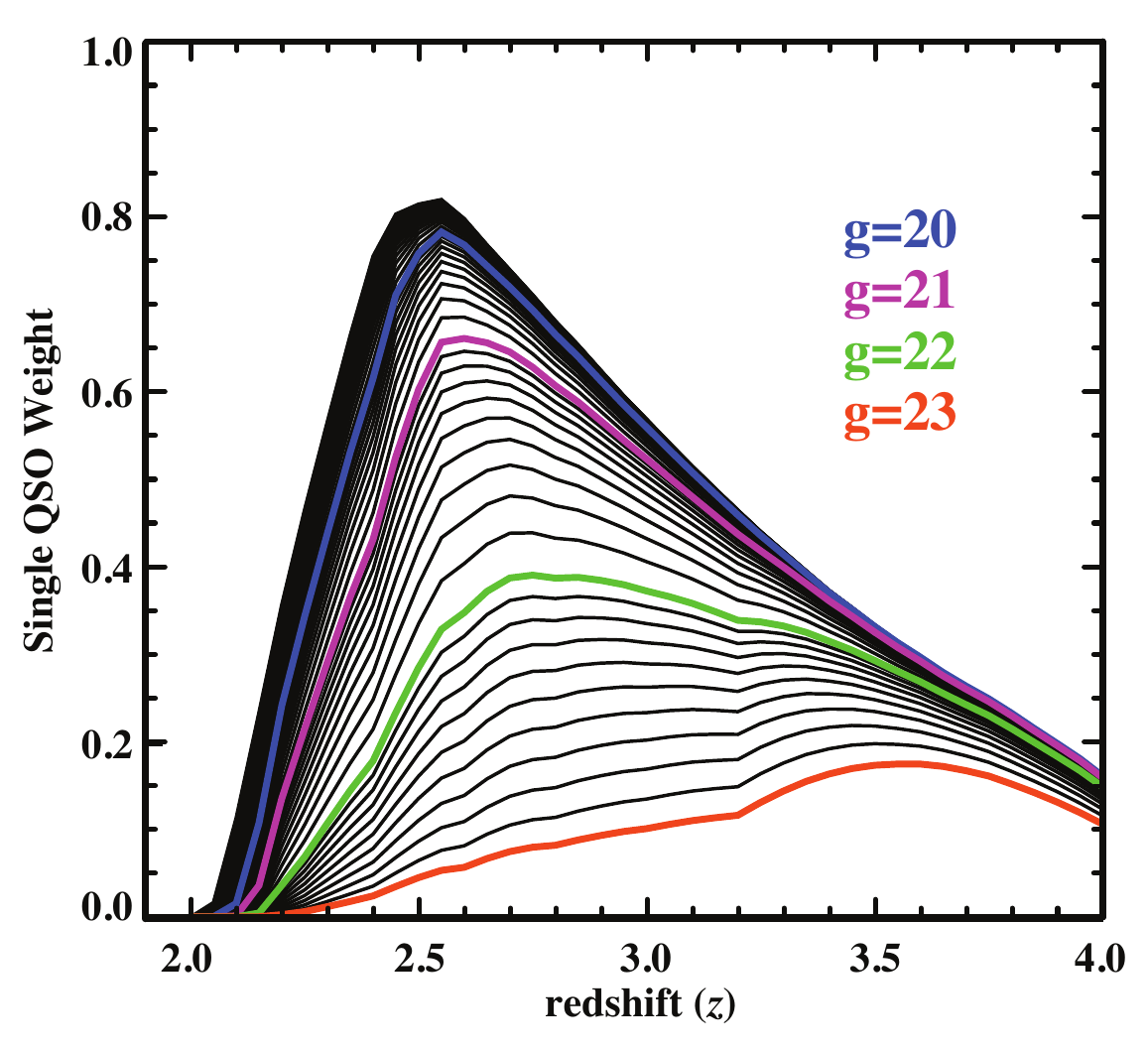}
	\caption[]
      {The McDonald Weight (or effectiveness as a dark energy BAO probe) of a QSO as a function of magnitude and redshift.  The lines are at 0.1 magnitude intervals. Brighter quasars have a higher weight, and so do quasars centered around a $z\sim2.5$.  For a detailed table of the numbers in this plot, see Table\;(\ref{tab:weights}).}\vspace{-0.6cm}
\label{fig:McDonaldWeights}
\end{center}
\end{figure}

We tested adjusting the likelihood method in this manner with weights ($w_{O'}$) calculated by Pat McDonald (private communication, see Fig.~\ref{fig:McDonaldWeights}, and Table~\ref{tab:weights}).   Here the weight is determined by the contribution of the quasar's Ly$\alpha$F to the BAO signal; a higher weight yields a higher signal-to-noise BAO measurement.  

The weight is a functional derivative of the overall BAO distance error squared with respect to the luminosity function. It can therefore be integrated over any achieved luminosity function (or summed over a set of quasars) to produce an estimate proportional to the BAO distance error squared that one would expect to achieve from that data set. There are two relevant factors affecting the value of a quasar: the noise level in the spectrum, and the density of quasars at a given redshift.  The low redshift cutoff comes primarily from the degradation in the signal-to-noise ratio at the blue end of the spectrograph, while the high-$z$ tail-off comes from the diminishing density of quasars with which to perform a cross-correlation for Ly$\alpha$F calculations. 

While using these weights did recover QSOs with a higher on average BAO signal, as expected fewer total quasars were recovered using this scheme, see Fig.\;(\ref{fig:weightingFigs}).  Ultimately it was decided to optimize the number of recovered quasars rather than the BAO signal for BOSS target selection.  Therefore this weighting scheme was not used in the final likelihood targeting algorithm.  However, depending on the goals of the user, a weighting scheme could be useful for future targeting purposes.

\subsection{Likelihood and XDQSO}
\label{subsec:XDQSO}

The likelihood method inspired a similar targeting approach, extreme-deconvolution quasar targeting \citep[XDQSO;][]{Bovy2011}. The training sets used in XDQSO are almost identical to the $QSO$ and $EE$ Catalogs used in the likelihood method.  XDQSO uses an extreme-deconvolution fit to these catalogs, such that they are represented by a small set of Gaussian distributions instead of large set of discrete objects.  Likelihood calculates probabilities as a straightforward sum over all objects, whereas XDQSO does a more algorithmically complicated fit.  However, once the catalog Gaussians are determined, XDQSO probabilities are much faster to calculate than those in the likelihood method.  In the limit where the $QSO$ and $EE$ Catalogs are extremely large, and their photometric errors are very small, the two algorithms are essentially equivalent.  However, for small, noisy training sets, likelihood represents the continuous color-distribution of quasars as discrete delta functions in color-space which can produce noisier results.  A study of the performance of the two methods shows that they are comparable in BOSS targeting efficiency, with XDQSO finding $\sim 1$ additional QSO deg$^{-2}$ at a given threshold \citep{Bovy2011}.  The similarity in targeting success is because the training sets used in likelihood are large and use high signal-to-noise data.

\subsection{Summary and Conclusions}
\label{subsec:bosstargeting}

In this paper we:
\begin{itemize}
\item Developed a new method for quasar target selection using photometric fluxes and a Bayesian probabilistic approach;
\item Demonstrated that this leads to the recovery of 15.9 ($z>2.2$) quasars deg$^{-2}$ from the SDSS Stripe 82 dataset when targeting at 40 targets deg$^{-2}$, with a completeness of 65\% and efficiency of 40\%;
\item Showed that the likelihood method recovers twice as many quasars as traditional ``color-box'' selection;
\item Tested for the effects of changing the input QSO catalog, using different luminosity functions and adding a weighting scheme to the likelihood calculations.
\end{itemize}

The likelihood method can easily be extended to include other attributes ($\mathbf{a}$) in addition to the photometric fluxes in the exponentials in Eq.\;(\ref{likeQSO}) and Eq.\;(\ref{likeEE}).  In addition, variability information, which has already been demonstrated as useful in quasar target selection \citep{Schmidt10, MacLeod2011, PD11}, could be incorporated.  Similarly, extending this method to include more color filters could help with target selection, as shown by \cite{Richards09_2}.

After a commissioning period in September-November 2009, the QSO targeting fibers were dividing into a uniformly selected CORE sample and a non-uniformly selected BONUS sample \citep{Ross11}.  The likelihood method, using the Jiang luminosity function, was used for targeting the CORE sample (20 targets deg$^{-2}$) for the first year of BOSS data taking.  The rest of the fibers (BONUS sample) were targeted by a combination of the output of the likelihood, KDE and NN methods using a neural network.  This approach allows us to combine both different methods and different photometric catalogs (SDSS, UKIDSS, GALEX) in the BONUS selection.  

After the first year of data taking, the CORE sample targeting switched to using the XDQSO method \citep{Bovy2011} and likelihood was then used in the BONUS sample as well as one of the inputs to the NN.  This switch was made because XDQSO performed slightly better at recovering high-redshift quasars and the priority of the target selection team was to maximize number of selected quasars.  We plan to release the probabilities from Eq.\;(\ref{probability}) in the project data releases of SDSS data.

\section*{Acknowledgments}

Thank you to Pat McDonald, Jo Bovy, Michael Strauss, Yolanda Hagar Slichter, and Adam Pauls for their contributions to this paper.

Funding for SDSS-III has been provided by the Alfred P. Sloan Foundation, the Participating Institutions, the National Science Foundation, and the U.S. Department of Energy Office of Science. The SDSS-III web site is http://www.sdss3.org/.

SDSS-III is managed by the Astrophysical Research Consortium for the Participating Institutions of the SDSS-III Collaboration including the University of Arizona, the Brazilian Participation Group, Brookhaven National Laboratory, University of Cambridge, University of Florida, the French Participation Group, the German Participation Group, the Instituto de Astrofisica de Canarias, the Michigan State/Notre Dame/JINA Participation Group, Johns Hopkins University, Lawrence Berkeley National Laboratory, Max Planck Institute for Astrophysics, New Mexico State University, New York University, Ohio State University, Pennsylvania State University, University of Portsmouth, Princeton University, the Spanish Participation Group, University of Tokyo, University of Utah, Vanderbilt University, University of Virginia, University of Washington, and Yale University.

This work was supported by the Director, Office of Science, Office of High Energy Physics, of the U.S. Department of Energy under Contract No. DE-AC02-05CH11231

\clearpage
\bibliographystyle{apj} 
\bibliography{biblio}

\begin{thebibliography}{46}
\expandafter\ifx\csname natexlab\endcsname\relax\def\natexlab#1{#1}\fi

\bibitem[{{Abazajian} {et~al.}(2009)}]{Abazajian09}
{Abazajian}, K.~N. {et~al.} 2009, \apjs, 182, 543

\bibitem[{{Adelman-McCarthy} {et~al.}(2008)}]{Adelman-McCarthy08}
{Adelman-McCarthy}, J.~K. {et~al.} 2008, \apjs, 175, 297

\bibitem[{{Aihara} {et~al.}(2011)}]{SDSSDR82011}
{Aihara}, H. {et~al.} 2011, \apjs, 193, 29

\bibitem[{{Albrecht} {et~al.}(2006)}]{DETF2006}
{Albrecht}, A. {et~al.} 2006, ArXiv Astrophysics e-prints

\bibitem[{{Bovy} {et~al.}(2011)}]{Bovy2011}
{Bovy}, J. {et~al.} 2011, The Astrophysical Journal, 729, 141

\bibitem[{{Croom} {et~al.}(2004)}]{Croom04}
{Croom}, S.~M. {et~al.} 2004, \mnras, 349, 1397

\bibitem[{{Croom} {et~al.}(2009)}]{Croom2009}
---. 2009, \mnras, 392, 19

\bibitem[{{Eisenstein} {et~al.}(2011){Eisenstein}, , {et~al.}}]{Eisenstein11}
{Eisenstein}, D.~J., , {et~al.} 2011, ArXiv e-prints

\bibitem[{{Fan}(1999)}]{Fan99}
{Fan}, X. 1999, \aj, 117, 2528

\bibitem[{{Fukugita} {et~al.}(1996){Fukugita}, {Ichikawa}, {Gunn}, {Doi},
  {Shimasaku}, \& {Schneider}}]{Fukugita1996}
{Fukugita}, M., {Ichikawa}, T., {Gunn}, J.~E., {Doi}, M., {Shimasaku}, K., \&
  {Schneider}, D.~P. 1996, \aj, 111, 1748

\bibitem[{{Gunn} {et~al.}(1998)}]{Gunn1998}
{Gunn}, J.~E. {et~al.} 1998, \aj, 116, 3040

\bibitem[{{Gunn} {et~al.}(2006)}]{Gunn2006}
---. 2006, \aj, 131, 2332

\bibitem[{{Hennawi} {et~al.}(2010)}]{Hennawi2010}
{Hennawi}, J.~F. {et~al.} 2010, \apj, 719, 1672

\bibitem[{{Hogg} {et~al.}(2001){Hogg}, {Finkbeiner}, {Schlegel}, \&
  {Gunn}}]{Hogg01}
{Hogg}, D.~W., {Finkbeiner}, D.~P., {Schlegel}, D.~J., \& {Gunn}, J.~E. 2001,
  \aj, 122, 2129

\bibitem[{{Hopkins} {et~al.}(2007){Hopkins}, {Richards}, \&
  {Hernquist}}]{HRH07}
{Hopkins}, P.~F., {Richards}, G.~T., \& {Hernquist}, L. 2007, The Astrophysical
  Journal, 654, 731

\bibitem[{{Ivezi{\'c}} {et~al.}(2004)}]{Ivezic04}
{Ivezi{\'c}}, {\v Z}. {et~al.} 2004, Astronomische Nachrichten, 325, 583

\bibitem[{{Jiang} {et~al.}(2006)}]{Jiang2006}
{Jiang}, L. {et~al.} 2006, \aj, 131, 2788

\bibitem[{{Lawrence} {et~al.}(2007)}]{Lawrence2007}
{Lawrence}, A. {et~al.} 2007, \mnras, 379, 1599

\bibitem[{{Lupton} {et~al.}(2001){Lupton}, {Gunn}, {Ivezi{\'c}}, {Knapp}, \&
  {Kent}}]{Lupton01}
{Lupton}, R., {Gunn}, J.~E., {Ivezi{\'c}}, Z., {Knapp}, G.~R., \& {Kent}, S.
  2001, in Astronomical Society of the Pacific Conference Series, Vol. 238,
  Astronomical Data Analysis Software and Systems X, ed. {F.~R.~Harnden Jr.,
  F.~A.~Primini, \& H.~E.~Payne}, 269

\bibitem[{{MacLeod} {et~al.}(2011)}]{MacLeod2011}
{MacLeod}, C.~L. {et~al.} 2011, \apj, 728, 26

\bibitem[{{Martin} {et~al.}(2005)}]{Martin2005}
{Martin}, D.~C. {et~al.} 2005, \apjl, 619, L1

\bibitem[{{Mortlock} {et~al.}(2011)}]{Mortlock11}
{Mortlock}, D.~J. {et~al.} 2011, ArXiv e-prints

\bibitem[{{Padmanabhan} {et~al.}(2008)}]{Padmanabhan08a}
{Padmanabhan}, N. {et~al.} 2008, \apj, 674, 1217

\bibitem[{{Palanque-Delabrouille} {et~al.}(2011)}]{PD11}
{Palanque-Delabrouille}, N. {et~al.} 2011, \aap, 530, A122+

\bibitem[{{Pier} {et~al.}(2003)}]{Pier03}
{Pier}, J.~R. {et~al.} 2003, \aj, 125, 1559

\bibitem[{{Richards} {et~al.}(2002)}]{Richards02}
{Richards}, G.~T. {et~al.} 2002, \aj, 123, 2945

\bibitem[{{Richards} {et~al.}(2004)}]{Richards2004}
---. 2004, The Astrophysical Journal Supplement Series, 155, 257

\bibitem[{{Richards} {et~al.}(2006)}]{Richards06}
---. 2006, The Astronomical Journal, 131, 2766

\bibitem[{{Richards} {et~al.}(2009)}]{Richards09_2}
---. 2009, \aj, 137, 3884

\bibitem[{{Ross} {et~al.}(2011{\natexlab{a}})}]{Ross11}
{Ross}, N.~P. {et~al.} 2011{\natexlab{a}}, ArXiv e-prints

\bibitem[{{Ross} {et~al.}(2011{\natexlab{b}})}]{Ross11DR9}
---. 2011{\natexlab{b}}, ArXiv e-prints

\bibitem[{{Sandage}(1965)}]{Sandage1965}
{Sandage}, A. 1965, \apj, 141, 1560

\bibitem[{{Schlegel} {et~al.}(2009){Schlegel}, {White}, \&
  {Eisenstein}}]{Schlegel2009}
{Schlegel}, D., {White}, M., \& {Eisenstein}, D. 2009, in ArXiv Astrophysics
  e-prints, Vol. 2010, astro2010: The Astronomy and Astrophysics Decadal
  Survey, 314--+

\bibitem[{{Schlegel} {et~al.}(1998){Schlegel}, {Finkbeiner}, \&
  {Davis}}]{SFD98}
{Schlegel}, D.~J., {Finkbeiner}, D.~P., \& {Davis}, M. 1998, The Astrophysical
  Journal, 500, 525

\bibitem[{{Schlegel} {et~al.}(2007)}]{Schlegel2007}
{Schlegel}, D.~J. {et~al.} 2007, in Bulletin of the American Astronomical
  Society, Vol.~38, American Astronomical Society Meeting Abstracts, 132.29--+

\bibitem[{{Schmidt} {et~al.}(2010){Schmidt}, {Marshall}, {Rix}, {Jester},
  {Hennawi}, \& {Dobler}}]{Schmidt10}
{Schmidt}, K.~B., {Marshall}, P.~J., {Rix}, H.~f., {Jester}, S., {Hennawi},
  J.~F., \& {Dobler}, G. 2010, \apj, 714, 1194

\bibitem[{{Schneider} {et~al.}(2007)}]{SDSSDR5}
{Schneider}, D.~P. {et~al.} 2007, \aj, 134, 102

\bibitem[{{Schneider} {et~al.}(2010)}]{Schneider10}
---. 2010, \aj, 139, 2360

\bibitem[{Sivia \& Skilling(2006)}]{Sivia2006}
Sivia, D. \& Skilling, J. 2006, Data analysis: a Bayesian tutorial, Oxford
  science publications (Oxford University Press)

\bibitem[{{Slosar} {et~al.}(2011)}]{Slosar2011}
{Slosar}, A. {et~al.} 2011, \apj, in prep.

\bibitem[{{Smith} {et~al.}(2002)}]{Smith02}
{Smith}, J.~A. {et~al.} 2002, \aj, 123, 2121

\bibitem[{{Stoughton} {et~al.}(2002)}]{Stoughton02}
{Stoughton}, C. {et~al.} 2002, \aj, 123, 485

\bibitem[{{Tucker} {et~al.}(2006)}]{Tucker06}
{Tucker}, D.~L. {et~al.} 2006, Astronomische Nachrichten, 327, 821

\bibitem[{{Y{\`e}che} {et~al.}(2009)}]{Yeche2009}
{Y{\`e}che}, C. {et~al.} 2009, ArXiv e-prints

\bibitem[{{Y{\`e}che} {et~al.}(2010)}]{Yeche10}
---. 2010, \aap, 523, A14+

\bibitem[{{York} {et~al.}(2000)}]{York2000}
{York}, D.~G. {et~al.} 2000, \aj, 120, 1579

\end{thebibliography}

\end{document}